\documentclass[fleqn,usenatbib]{mnras}

\usepackage{newtxtext,newtxmath}

\usepackage[T1]{fontenc}

\DeclareRobustCommand{\VAN}[3]{#2}
\let\VANthebibliography\thebibliography
\def\thebibliography{\DeclareRobustCommand{\VAN}[3]{##3}\VANthebibliography}

\usepackage{graphicx}	
\usepackage{amsmath}	
\usepackage{xcolor}    

\title[Composite model of a lensed quasar]{Disentangling the dark and stellar mass through precise lens modelling of the JWST observation of lensed quasar WFI2033--4723}

\author[Tian Li et al.]{
Tian Li,$^{1}$\thanks{E-mail: tian.li@port.ac.uk}
Thomas E. Collett$^{1}$,
Coleman M. Krawczyk$^{1}$,
Wolfgang J. R. Enzi$^{1}$,
Aymeric Galan$^{2}$
\\
$^{1}$Institute of Cosmology and Gravitation, University of Portsmouth, Burnaby Rd, Portsmouth PO1 3FX, UK\\
$^{2}$Department of Astronomy, University of Geneva, Chemin d’Ecogia 16, 1290 Versoix, Switzerland \label{unige}
}

\date{Accepted XXX. Received YYY; in original form ZZZ}

\pubyear{\the\year{}}

\begin{document}
\label{firstpage}
\pagerange{\pageref{firstpage}--\pageref{lastpage}}
\maketitle

\begin{abstract}
We use high-resolution JWST/NIRCam imaging and measured time delays to model the quadruply imaged quasar WFI2033--4723 with a composite stellar plus dark-matter mass model. We first construct an elliptical power-law baseline model and recover Fermat-potential differences (fpd) consistent with previous HST-based and JWST-based analyses, providing a reference scale for composite modelling. We then replace the total mass profile with a physically motivated decomposition in which the stellar mass follows a multi-Gaussian expansion of the lens light, with a free radial mass-to-light gradient, and the dark matter is described by a generalized Navarro--Frenk--White (gNFW) halo. Using two external cosmological priors, Planck+DESI and Pantheon+SH0ES, the measured time delays constrain the mass-sheet-transformation freedom that would otherwise damage the stellar--dark-matter decomposition. In both cosmological cases, the stellar normalization lies between the expectations for Chabrier and Salpeter initial mass functions, while the radial mass-to-light gradient is not strongly required by the data (mildly positive). The dark matter halo has an inner slope \(\gamma_{\rm in}\simeq1.3\), steeper than a standard NFW cusp, and the main astrophysical conclusions are insensitive to the adopted cosmological prior. This work shows that composite time-delay lens modelling can effectively separate baryons from dark matter. As a qualitative check, we reverse the logic and use our composite lens model without kinematic information to infer the cosmology instead. However, the strong degeneracy between \(H_0\) and the halo scale radius \(R_s\) prevents a robust standalone constraint.
\end{abstract}




\section{Introduction}

Dark matter dominates the mass of massive galaxies but cannot be seen directly, so its distribution must be inferred from its gravitational influence on luminous tracers. The two primary methods are galaxy kinematics \citep{Rubin1978, Rubin1980} and gravitational lensing \citep[see][for a review]{Treu2010review}. Strong lensing directly probes the mass distribution: when a foreground galaxy aligns with a background source, it bends the source’s light into multiple images or an Einstein ring. Because lensing traces the total gravitational potential, it constrains the combined distribution of luminous and dark matter.

For a galaxy-scale strong lens, the projected mass enclosed within the Einstein radius can be measured at the $\sim 1\%$ level, and this constraint is largely insensitive to the detailed functional form of the mass distribution \citep{Schneider1992}. The precision, however, is local, because lensing mostly probes the potential in the annulus spanned by the multiple images \citep{Birrer2021,O'Riordan2020}. A precise enclosed-mass measurement can therefore still be consistent with a range of radial profiles and stellar--dark-matter decompositions, unless additional assumptions or observables are introduced.

In the cold dark matter (CDM) paradigm, dark matter haloes are expected to have NFW-like profiles \citep{NFW1997}, while the stellar component of massive galaxies is strongly centrally concentrated. The combination of these two components is close to a power law over the radial range probed by galaxy-scale strong lenses, a behaviour known as the bulge--halo conspiracy \citep{True2004,Koopmans2006,Koopmans2009,Auger2010,Dutton2014}. This motivates the elliptical power-law (EPL) profile as a simple and efficient baseline model for strong-lens analyses \citep{Shajib2021,Etherington2023,Tdcosmo2025}. However, to relate lensing constraints more directly to CDM expectations and to allow deviations from an exact power law, one can instead use composite models in which the stellar and dark matter components are modelled separately \citep{Sonnenfeld2012,Oldham2018,Collett2018,Melo2025, Tian2026}. In such models, the dark matter is often represented by a generalised NFW (gNFW) profile whose inner slope $\gamma_{\rm DM}$ is allowed to vary \citep{Keeton2001,Wyithe2001}.

The inner slope of galaxy dark matter haloes carries information about both collisionless CDM and baryonic physics. Cooling and in-situ star formation can deepen the central potential and contract the halo \citep{Blumenthal1986,Gnedin2004}, whereas bursty feedback can flatten the inner profile by driving rapid potential fluctuations \citep{PontzenGovernato2012,Martizzi2013}. The observed slope is therefore expected to depend on the galaxy mass and assembly history, as also seen in simulations \citep{DiCintio2014a,Tollet2016,Dutton2016}.

\begin{figure*}
    \centering
    \includegraphics[width=1\linewidth]{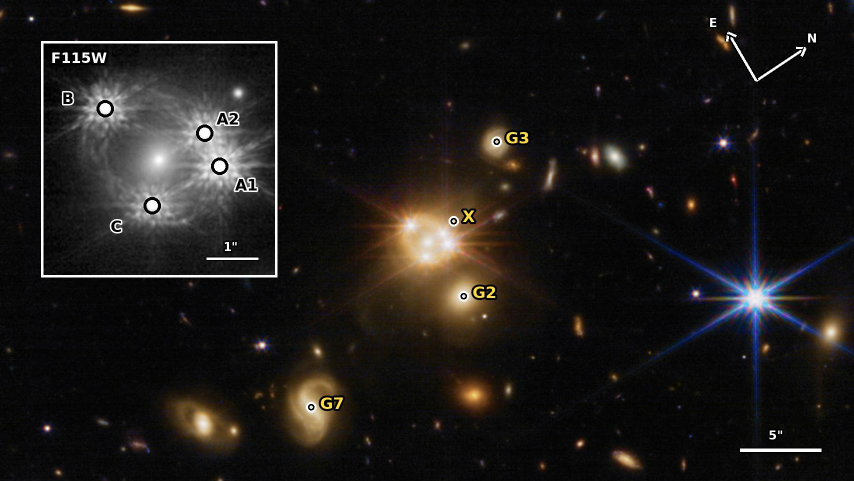}
    \vspace{-0.4cm}
    \caption{Colour composite of the WFI2033 field constructed from four JWST/NIRCam bands: F115W, F150W, F277W, and F356W. The top left image shows a zoomed F115W image of the lens system, with the four quasar images labelled as B, C, A1, and A2. The nearby galaxies included in the lens model are marked as satellite X and perturbers G2, G3, and G7. The compass indicates north and east. The scale bars correspond to \(1^{\prime\prime}\) in the zoomed inset and \(5^{\prime\prime}\) in the full-field image. A full-resolution version is available online.\protect\footnotemark}
    \vspace{-0.5cm}
    \label{fig:colorimg}
\end{figure*}
\footnotetext{\url{https://tianliscope.pages.dev/photo?album=spacetelescope&id=wfi2033}}

Disentangling the stellar and dark matter components with lens modelling, and hence constraining $\gamma_{\rm DM}$ on kiloparsec scales, is challenging because the stellar contribution depends on the stellar mass-to-light ratio ($M/L$). This quantity is sensitive to the assumed initial mass function (IMF), and observations of early-type galaxies (ETGs) suggest that the IMF can be bottom-heavy in their central regions \citep{Auger2010}. Radial IMF variations, and the associated $M/L$ gradients, may then become progressively more Milky Way-like beyond the inner few kiloparsecs \citep{vanDokkum2017,Sarzi2018,Sonnenfeld2018,Collett2018,Mehrgan2024}.

Another challenge is the mass-sheet transformation (MST), a fundamental degeneracy of lens modelling in which a rescaling of the convergence and source plane leaves the lensing observables unchanged \citep{Falco1985,Schneider2013}. In a composite model, changes in the stellar \(M/L\) gradient and dark-matter profile can generate an MST-like degeneracy, even if they are not mathematically identical to a pure external mass sheet \citep{Li2025a}. Changing the inner slope of the gNFW halo, for example, can make the total profile more concave or convex while preserving a similar lensing fit.

A common way to add information that helps break this degeneracy is to combine strong lensing with stellar dynamics. Population analyses of early-type strong lenses have generally found dark matter haloes close to NFW, with results that are broadly robust to plausible stellar $M/L$ gradients \citep{Shajib2021,Sheu2025}. Individual composite analyses, however, have reached mixed conclusions: some studies inferred very cuspy haloes \citep{Sonnenfeld2012,Oldham2018}, while others found profiles closer to NFW \citep{Collett2018,Melo2025}. These inferences remain sensitive to assumptions about the lens light, stellar $M/L$, halo geometry, and stellar dynamics, including anisotropy and projection effects \citep{Huang2025}.

A complementary way to constrain the MST is to use the geometry of double-source-plane lenses (DSPLs): \citet{Tian2026} showed with the Jackpot lens that the additional source plane can help break the MST and constrain a composite stellar-plus-dark-matter model. However, this approach introduces its own systematic uncertainty because the mass profile of the intermediate source must be modelled and can be degenerate with the main-lens mass profile.

Time delays offer an independent constraint on the same MST-like freedom. Lensed quasars with well-measured time delays have long been used to infer the Hubble constant \citep[see][and references therein]{Tdcosmo2025}, with the MST being one of the dominant systematic uncertainties. Here we reverse this logic. If the background cosmology is assumed or externally constrained, the observed time delays between the quasar images can instead be used to constrain the MST, and therefore to distinguish between different stellar-plus-dark-matter decompositions that would otherwise fit the imaging data equally well. This is similar in spirit to \citet{DobkeKing2006}, who used an external \(H_0\) prior and time-delay lenses to constrain the slope of a power-law density profile, but here we apply the idea to a more flexible composite mass model.

In this work, we apply this strategy to the quadruply imaged quasar WFI2033--4723 using high-resolution JWST/NIRCam imaging and the measured time delays. The quasar source has spectroscopic redshift \(z_{\rm s}=1.662\) \citep{sluse2012}, and the main deflector has spectroscopic redshift \(z_{\rm d}=0.6575\) from the H0LiCOW VLT/MUSE survey \citep{sluse2019}; the spectroscopic redshift uncertainties are negligible for the lens-modelling uncertainties considered here ($\sigma_{z_{\rm d}}=2\times10^{-4}$). WFI2033--4723 was modelled by H0LiCOW with both power-law and composite mass profiles, yielding \(H_0=71.6^{+3.8}_{-4.9}\,{\rm km\,s^{-1}\,Mpc^{-1}}\) from HST imaging \citep{rusu2020}. More recently, TDCOSMO XX re-modelled the system with JWST/NIRCam data and found results consistent with the HST-based analysis \citep{Williams2025TDCOSMOXX}. These results make WFI2033--4723 a useful test case for whether the same data can constrain a more physical stellar-plus-dark-matter decomposition.

We first construct an EPL baseline model to obtain a precise reference lens solution and Fermat-potential differences (fpd), and then replace the total mass profile using the star-plus-dark-matter framework introduced by \citet{Tian2026}. In this composite model, the stellar mass follows a flexible multi-Gaussian expansion (MGE) of the lens light and the dark matter is described by an elliptical gNFW halo. We allow the stellar normalisation and radial stellar \(M/L\) gradient to vary, so that the model can explore the same stellar--dark-matter trade-off that underlies the MST. We combine the imaging constraints with the measured quasar time delays under externally motivated cosmological assumptions to test whether the time-delay information can disentangle the stellar and dark matter contributions without relying primarily on stellar kinematics.

We then specify the external cosmological information used to set the time-delay distance scale. We analyse the composite models under two representative cosmological priors in a flat \(\Lambda\)CDM cosmology: a local-distance-ladder prior from Pantheon+ combined with SH0ES, \(H_0=73.50\pm1.07\,{\rm km\,s^{-1}\,Mpc^{-1}}\) \citep{Brout2022PantheonPlus,Riess2022SH0ES}, and an early-Universe/large-scale-structure prior from Planck CMB measurements combined with DESI BAO constraints, \(H_0=68.17\pm0.28\,{\rm km\,s^{-1}\,Mpc^{-1}}\) \citep{Planck2018,DESI2025BAO}. Comparing these two cases allows us to assess whether the inferred stellar \(M/L\) normalisation, radial \(M/L\) gradient, and gNFW inner slope are sensitive to the adopted cosmological prior, or are instead driven mainly by the JWST imaging.

\section{Data}
\label{sec:data}

Our analysis uses \textsc{JWST}/NIRCam observations of WFI2033--4723 from GTO 1198 (PI: Stiavelli). The data set contains four NIRCam filters, F115W, F150W, F277W, and F356W. In this work, we mainly model  the reduced F115W image, which provides high angular resolution and the main constraints from the quasar images and lensed host-galaxy arcs. The modelling uses a \(150\times150\)-pixel cutout, corresponding to approximately \(4.6^{\prime\prime}\) on a side, with a drizzled pixel scale of \(0.0307^{\prime\prime}\) per pixel. At the lens redshift, F115W corresponds to \(\lambda_{\rm rest}\simeq0.70\,\mu{\rm m}\), close to the rest-frame wavelength of the \textit{HST}/ACS F814W imaging used for the Jackpot lens at \(z=0.222\) in \citet{Tian2026}; this makes the stellar-light based mass decomposition in the two analyses broadly comparable.

WFI2033--4723 is a quadruply imaged quasar first reported by \citet{morgan2004}. We use the COSMOGRAIL time-delay measurements adopted by H0LiCOW for this system \citep{bonvin2019,rusu2020}: \(\Delta t_{\rm B-A1}=-36.2^{+1.6}_{-2.3}\) days, \(\Delta t_{\rm B-A2}=-37.3^{+2.6}_{-3.0}\) days, and \(\Delta t_{\rm B-C}=-59.4\pm1.3\) days. The local environment includes the nearby satellite X and the galaxy perturbers G2, G3, and G7, which were identified in previous time-delay cosmography analyses as perturbers capable of producing non-negligible shifts in the inferred \(H_0\) \citep{sluse2019,rusu2020}.

\section{Lens Modelling}
\subsection{Lensing theory}
We use the following lensing conventions throughout the analysis.

For a lensed image at position $\boldsymbol{\theta}$, the scaled deflection angle of a lens galaxy $\boldsymbol{\alpha}(\boldsymbol{\theta})$ is related to its lensing potential $\psi$ via
\begin{equation}
\boldsymbol{\alpha}(\boldsymbol{\theta})=\nabla \psi(\boldsymbol{\theta})\,,
\end{equation}
and the relation between lensing potential and lensing convergence is
\begin{equation}
\kappa(\boldsymbol{\theta})=\frac{1}{2} \nabla^2 \psi(\boldsymbol{\theta})\,,
\end{equation}
where convergence is defined as
\begin{equation}
\kappa(\boldsymbol{\theta}) \equiv \frac{\Sigma(\boldsymbol{\theta})}{\Sigma_{\mathrm{cr}}}\,.
\end{equation}
The convergence is the lens surface mass density normalised by the critical lensing surface density,
\begin{equation}
\Sigma_{\mathrm{cr}} \equiv \frac{c^2 D_{\mathrm{s}}}{4 \pi G D_{\mathrm{l}} D_{\mathrm{ls}}}\,,
\end{equation}
where \(D\) is the angular diameter distance between two objects, and the subscripts l and s denote the lens galaxy and the source galaxy, respectively.

The time delay between two images A and B is given by:
\begin{equation}
\Delta t_{\mathrm{AB}}=\frac{D_{\Delta t}}{c}\left(\phi\left(\theta_{\mathrm{A}}, \beta\right)-\phi\left(\theta_{\mathrm{B}}, \beta\right)\right)
\end{equation}
where $\phi(\theta, \beta)$ is the Fermat potential:
\begin{equation}
\phi(\theta, \beta)=\left[\frac{(\theta-\beta)^2}{2}-\psi(\theta)\right]
\end{equation}
and the time-delay distance $D_{\Delta t}$ is:
\begin{equation}
D_{\Delta t} \equiv\left(1+z_{\mathrm{l}}\right) \frac{D_{\mathrm{l}} D_{\mathrm{s}}}{D_{\mathrm{ls}}}
\end{equation}
Time-delay cosmography uses measured time delays with mass profiles from lens modeling and kinematics to constrain the time-delay distance, hence constraining the cosmological parameters. The Hubble constant is inversely proportional to the time-delay distance:
\begin{equation}
H_0 \propto D_{\Delta t}^{-1} .
\end{equation}

\subsection{Mass-sheet transform}
Constraining the mass profile of a lens galaxy is challenging due to the MST \citep{Falco1985}. The MST is defined by the equation:
\begin{equation}
\lambda \beta=\theta-\lambda \alpha(\theta)-(1-\lambda) \theta
\end{equation}
This transform is a multiplicative modification of the lens equation, preserving the image positions under a linear displacement of the source, $\beta \rightarrow \lambda \beta$. The term $(1 - \lambda)$ represents an infinite sheet of mass. Observables that are sensitive to the absolute source size, intrinsic magnification, or lensing potential can help break this degeneracy. 

The MST on convergence is:
\begin{equation}
\label{eq:mst}
    \kappa_{\lambda}(\theta) = \lambda \kappa(\theta) + 1 - \lambda
\end{equation}
where $\kappa_\lambda$ is the convergence after the mass-sheet transform. This equation indicates that if we scale the convergence by $\lambda$ and add a convergence sheet with $\kappa=1-\lambda$, we would observe the same image as with the original convergence.

\section{Lens modelling strategy}

\begin{table}
\centering
\small
\renewcommand{\arraystretch}{1.0}
\setlength{\tabcolsep}{4pt}
\begin{tabular}{@{}lcc@{}}
\hline
\textbf{Parameter}                         & \textbf{EPL baseline}                    & \textbf{star+gNFW} \\
\hline
$\theta_{\rm E}$                           & $\mathcal{U}(0.9,\,1.1)$                  & -- \\
$\gamma_{\rm EPL}$                         & $\mathcal{U}(1.2,\,2.8)$                  & -- \\
$(e_1,e_2)$                                & $\mathrm{TN}(0,\,0.25;\,-0.5,\,0.5)$      & \(\mathrm{TN}(0,\,0.25;\,-0.5,\,0.5)\) \\
$(x_{\rm c},y_{\rm c})$                    & $\mathrm{TN}(0,\,0.1;\,-0.5,\,0.5)$       & $\mathrm{TN}(0,\,1;\,-0.4,\,0.4)$ \\
$\Upsilon_{\kappa}$                        & --                                        & $\mathcal{U}(0,\,5)$ \\
$\nabla M/L$                               & --                                        & $\mathcal{U}(-0.6,\,0.6)$ \\
$\kappa_{s,halo}$                                 & --                                        & $\mathcal{U}(0,\,1)$ \\
$\gamma_{\rm in}$                          & --                                        & $\mathcal{U}(0.6,\,2.0)$ \\
$R_s$                                      & --                                        & $\mathcal{U}(2,\,20)$ \\
$(\gamma_1,\gamma_2)$                      & $\mathcal{U}(-0.5,\,0.5)$                 & $\mathcal{U}(-0.5,\,0.5)$ \\
$\theta_{\rm E}^{\rm X}$                & $\mathcal{U}(0,\,0.25)$                  & $\mathcal{U}(0,\,0.25)$ \\
$\theta_{\rm E}^{\rm G2}$                  & $\mathcal{U}(0.5,\,0.7)$                 & $\mathcal{U}(0.5,\,0.7)$ \\
$\theta_{\rm E}^{\rm G3,G7}$               & scaled from \(\theta_{\rm E}^{\rm G2}\)  & scaled from \(\theta_{\rm E}^{\rm G2}\) \\
$I_{\rm lens}$                             & inner 3 fixed; outer 2 free & fixed EPL light \\
$\kappa_{\rm ext}$                         & --                                        & \(0.059^{+0.077}_{-0.047}\) \\
$(H_0,\Omega_{\rm m})_{\rm P+SH0ES}$       & --                                        & \(H_0=73.50\pm1.07\) \\
$(H_0,\Omega_{\rm m})_{\rm Planck+DESI}$   & --                                        & \(H_0=68.17\pm0.28\) \\
\hline
\end{tabular}
\caption{Priors used for the EPL baseline and the star+gNFW time-delay model. \(\mathcal{U}\) and \(\mathrm{TN}\) denote uniform and truncated Gaussian priors. The singular isothermal sphere (SIS) perturber priors are shared by both mass models. Following the notation of \citet{Tian2026}, \(\Upsilon_{\kappa}\) denotes the overall scale applied to the normalized stellar MGE in convergence. For the cosmological priors, the table lists only the marginal \(H_0\) constraints for compactness, while the inference uses the corresponding two-dimensional posterior constraints in the \((H_0,\Omega_{\rm m})\) plane as priors. The inner three MGE light components are fixed during the Hamiltonian Monte Carlo (HMC) sampling stage and the outer two MGE light components are free.}

\label{tab:prior_all}
\vspace{-0.5cm}
\end{table}

\label{sec:lensmodel_Setup}
We used the open-source lens modelling code {\tt Herculens}\footnote{\url{https://github.com/Herculens/herculens}} \citep{Galan2022_herculens}, which includes multiplane lensing capabilities (see, e.g. \citealt{Enzi2025}). {\tt Herculens} is built on the automatic differentiation and compilation features of JAX\footnote{\url{https://docs.jax.dev/en/latest/}} and can run on graphics processing units (GPUs). Automatic differentiation efficiently computes partial derivatives without manually deriving or implementing them \citep{Baydin2018}, allowing us to evaluate likelihood gradients with respect to all free parameters.

The modelling pipeline has four components: the light models for the lens and source, the point-spread function (PSF), the lens mass model, and the sampling strategy. We first initialise the lens model with an elliptical power-law mass profile and then transition to a composite model consisting of stars plus a generalized NFW dark-matter halo (star$+$gNFW).

\subsection{Light model}
\subsubsection{Lens light}
We model the lens light as a sum of multiple Gaussian components:
\begin{equation}
\label{eq:MGE_light}
F_{\mathrm{light}}(x,y) \;\approx\; \sum_{n=1}^{N}
A_{\mathrm{light},n}\,
\exp\!\left[-\frac{R^2}{2\,\sigma_n^{2}}\right],
\end{equation}
where, for each Gaussian component, the amplitude $A_{\mathrm{light},n}$, the axis ratio $q_n$ ($R^2 \equiv q_n\,(x-x_n)^2 + \frac{(y-y_n)^2}{q_n}$), the centroid $(x_n,y_n)$, and the ellipticity ($e_1, e_2$) are treated as independent free parameters. We adopt $N=5$ Gaussians with dispersions $\{\sigma_n\}$ uniformly spaced in $\log \sigma_n$ between $0.001$ and $3$~arcsec.

We parametrise the ellipticity using the Cartesian components $(e_1,e_2)$ instead of axis ratio $q$ and position angle $\phi$. This avoids the $\pi$-periodicity of $\phi$ and provides a continuous parameter space, which improves the efficiency and robustness of the inference. We adopt the standard mapping:
\begin{equation}
\label{eq:e1e2_def}
e_{1,n} = \frac{1-q_n}{1+q_n}\cos(2\phi_n), 
\qquad
e_{2,n} = \frac{1-q_n}{1+q_n}\sin(2\phi_n).
\end{equation}

\subsubsection{Source light parametric}

We first model the extended quasar host with a simple parametric surface-brightness profile to obtain a stable initial lens solution. WFI2033--4723 contains a single lensed background source in our model. We use one elliptical Gaussian component for the extended host galaxy, with free amplitude, centroid, scale radius, and ellipticity. The parametric source is not meant to reproduce the final host morphology; it only supplies an initial approximation to the arc light. More importantly, it provides a close starting point for the mass profile, which will later be refined when we switch to the pixelated source model described below.

\subsubsection{Source light pixelated}
We then model the sources as fields defined on regular Cartesian pixel grids that can be treated as a Gaussian process (GP). The source is painted on an adaptive source grid. At each sampling step, we trace the image-plane arc mask to the source plane using the current mass model, and choose the source grid as the smallest square enclosing all traced mask pixels.

The field value in each pixel forms an element of the vector $\vec{s}$. In Fourier space, the mode amplitudes can be determined by a power spectrum \citep[see, e.g., ][]{Galan2024, RustigEtAl2024_LensCharm}:
\[
\vec{s} \;=\; \mathcal{F}^{-1}\!\big[\,\sqrt{P}\,\odot\,\vec{\xi}\,\big],
\]
where \(P\) is the Matérn power spectrum evaluated on the Fourier grid, 
\(\odot\) denotes element-wise multiplication, and $\vec{\xi}$ is standard Gaussian white noise. The Matérn power spectrum \citep[see e.g.][]{Stein2012Interpolation} corresponding to a Matérn covariance kernel is
\begin{equation}
P(k)=\sigma^{2}\,4\pi n\left(\frac{2n}{\rho^{2}}\right)^{n}\left(\frac{2n}{\rho^{2}}+k^{2}\right)^{-(n+1)},
\end{equation}
where $\sigma$ sets the overall amplitude, $\rho$ is the correlation length, and $n$ controls the smoothness. We also fit for these parameters using a Jeffreys prior during the inference. 

Applying the discrete inverse Fourier transform \(\mathcal{F}^{-1}\) yields the pixelated source image \(\vec{s}\), which is a realisation of a Gaussian process with Matérn covariance.

To enforce non-negative source-plane surface brightness while keeping the model differentiable, we apply a smooth positivity map to the latent pixel field,
\begin{equation}
\vec{s}\;\longrightarrow\;\frac{\mathrm{softplus}(h\,\vec{s})}{h}, 
\mathrm{softplus}(x)=\log\!\left(1+e^{x}\right), h = 100.
\end{equation}
The softplus function is strictly positive and smooth: for \(x\ll 0\), \(\mathrm{softplus}(x)\approx e^{x}\) so it approaches \(0\) exponentially, while for \(x\gg 0\), \(\mathrm{softplus}(x)\approx x\) and thus becomes approximately linear. The scale parameter \(h\) controls how sharply the mapping transitions from near-zero values for negative inputs to an approximately identity mapping for positive inputs.

We use \(74\times74\)-pixel grids for the source. This grid size is chosen after the parametric modelling stage from the ray-tracing result of the parametric mass profile, and it provides sufficient ray-tracing samples per pixel to satisfy the Nyquist criterion: on average, four image-plane pixels map onto one source pixel.

\subsection{PSF modelling}

The PSF must be modelled carefully because the four quasar images are bright and lie close to the extended arcs. Even small PSF residuals can be absorbed by the lensed host-galaxy reconstruction or by the quasar amplitudes, and can therefore bias the lens model if they are not treated explicitly. Empirical studies of JWST/NIRCam imaging have shown that the PSF varies across the detector field and between exposures \citep{Nardiello2022JWSTPSF,Zhuang2024JWSTPSF}. These variations reflect both position-dependent optical/projection effects across the focal plane and exposure-to-exposure changes in the effective wavefront and sampling. TDCOSMO XX identified PSF modelling as a leading uncertainty for JWST lens modelling of WFI2033--4723 \citep{Williams2025TDCOSMOXX}. In WFI2033--4723, the quasar images are bright, have small formal statistical uncertainties, and are calibrated using field stars at other positions on the JWST focal plane, so the stellar PSF cannot perfectly reproduce the local PSF at the lens position.

\subsubsection{Initial PSF and PSF correction}

We first generate an initial PSF using \textsc{STPSF} for the F115W observation. To model the spatial variation of the PSF in the WFI2033--4723 field, we select a set of isolated field stars as empirical PSF constraints and fit them jointly with stochastic variational inference (SVI). The field-star model starts from the \textsc{STPSF} kernel and allows a flexible PSF correction, producing the modelled initial PSF used in the lens modelling. During the lens modelling itself, the lensed quasar images provide a second, on-the-fly PSF correction step. This allows the model to capture the local mismatch between the field-star PSF and the quasar-image PSF, including differences caused by position-dependent PSF variation and by the different spectral energy distributions of stars and quasars.

We parameterize the PSF correction as a positive multiplicative perturbation to a reference PSF. Let \(\mathrm{PSF}_0(\boldsymbol{x})\) be the reference PSF. We introduce a pixelized correction field \(\delta_{\rm PSF}(\boldsymbol{x})\) in log space and define the unnormalised corrected PSF as
\[
\mathrm{PSF}_{\rm corr}(\boldsymbol{x})
=
\mathrm{PSF}_0(\boldsymbol{x}) \exp\left[\delta_{\rm PSF}(\boldsymbol{x})\right].
\]
After this multiplicative correction, we renormalise \(\mathrm{PSF}_{\rm corr}\) so that its total flux is unity. This parameterization guarantees that the corrected PSF remains positive, while the normalisation preserves the PSF flux. It also has the useful property that \(\delta_{\rm PSF}=0\) corresponds exactly to the input PSF before renormalisation, so the correction is only applied where it is supported by the imaging data.

We assign independent Gaussian priors to the log-correction pixels,
\[
\delta_{\rm PSF}(\boldsymbol{x}) \sim \mathcal{N}(0,1).
\]

We prefer this log-multiplicative form over an additive correction because an additive perturbation can produce negative PSF values and requires additional constraints to preserve the total flux. The multiplicative correction instead describes relative deviations from the reference PSF, which is better matched to the expected PSF mismatch and leads to a more stable inference problem. This parameterization is also convenient for Hamiltonian Monte Carlo, since the sampled variables \(u(\boldsymbol{x}) \equiv \delta_{\rm PSF}(\boldsymbol{x})\) live in an unconstrained Euclidean space, while the deterministic transformation above maps them to a positive PSF. A correlated prior on the PSF-correction field, similar in spirit to the regularised PSF correction used in \textsc{STARRED} \citep{michalewicz2023,millon2024}, could reduce the risk of overfitting small-scale noise in the correction. We do not include this extension here.

\subsubsection{Extra PSF uncertainty}

The field-star modelling also provides an empirical estimate of the residual PSF mismatch. For each selected star, we compute the difference between the observed stellar cutout and the modelled initial PSF rendered at the fitted sub-pixel position and flux, after subtracting the fitted local background. We then align these residuals to a common PSF-centred frame and estimate a relative PSF-error map,
\begin{equation}
\alpha_{\rm PSF}^{2}(\boldsymbol{x})=\frac{\sum_j M_j^2(\boldsymbol{x})\left(R_j^2(\boldsymbol{x})-\sigma_j^2(\boldsymbol{x})\right)}{\sum_j M_j^4(\boldsymbol{x})},
\end{equation}
where \(R_j\), \(M_j\), and \(\sigma_j\) are the residual image, PSF-only model, and original error map of star \(j\) after alignment. In the lens-modelling sequence, we first fit the system without the pixelized PSF correction, then activate the PSF correction and use the resulting point source positions, fluxes, and corrected PSF to propagate \(\alpha_{\rm PSF}\) onto the quasar images. The final root-mean-square (RMS) map is
\begin{equation}
\sigma_{\rm tot}^{2}(\boldsymbol{x})=\sigma_{\rm img}^{2}(\boldsymbol{x})+\sum_i\left[\alpha_{{\rm PSF},i}(\boldsymbol{x})I_{{\rm PSF},i}(\boldsymbol{x})\right]^2,
\end{equation}
where \(I_{{\rm PSF},i}\) is the rendered model of quasar image \(i\), and \(\alpha_{{\rm PSF},i}\) is the aligned error map shifted to that image position. The resulting RMS map is used in the subsequent likelihood, so pixels most affected by PSF mismatch are down-weighted.

\subsection{Mass Model}

We first parametrise our mass model as an elliptical power-law mass model. Then we use a composite model of star plus dark matter which includes an MGE stellar mass component and a gNFW profile for the dark matter halo mass component approximated with a three-dimensional MGE of 20 Gaussians.

\subsubsection{Elliptical Power-Law}
Traditional strong-lens analyses often describe the total projected mass distribution with an EPL model. The convergence ($\kappa_{\mathrm{EPL}}$) of the EPL model can be parametrised as
\begin{equation}
\kappa_{\mathrm{EPL}}(R, \gamma, q)=\frac{3-\gamma}{2}\left(\frac{\theta_\mathrm{E}}{R}\right)^{\gamma-1}\,,
\end{equation}
where \(R\) is the elliptical radius, defined by \(R^2=q x'^2+y'^2/q\) in coordinates \((x',y')\) aligned with the mass major axis, \(\gamma\) is the logarithmic density slope and \(\theta_\mathrm{E}\) is the Einstein radius. For \(\gamma=2\), the model reduces to the singular isothermal ellipsoid (SIE), whose circular limit is the singular isothermal sphere (SIS). In our modelling, the masses associated with the nearby galaxies are modelled with SIS profiles.

\subsubsection{Stellar mass component}
Our fiducial model consists of stars plus dark matter. The stellar mass distribution is tied to the multi-Gaussian light model: the amplitude of each Gaussian is linked to its contribution to the stellar convergence, such that the relative convergence amplitudes between Gaussians are fixed by the light model. An $M/L$ ratio is required to map the light model to a mass model; here we treat the (global) $M/L$ as a free parameter. Following \citet{Oldham2018}, we also allow for a radial $M/L$ gradient:
\begin{equation}
\Upsilon_{\star}(\sigma_i) \propto \sigma_i^{\nabla M/L},
\end{equation}
where $\Upsilon_{\star}$ is the mass-to-light ratio assigned to each Gaussian component, $\sigma_i$ is the standard deviation of each Gaussian, and $\nabla M/L$ is the slope of the $M/L$ gradient. Here $\nabla M/L$ is an additional free parameter that controls a smooth, global radial variation of the stellar $M/L$ across the Gaussian components; $\nabla M/L=0$ corresponds to a constant $M/L$. A positive \(\nabla M/L\) means that \(M/L\) increases with radius, while a negative \(\nabla M/L\) means that \(M/L\) decreases with radius.
We normalise the stellar-mass MGE such that the Gaussian amplitudes sum to unity in convergence.  In practice, this means we do not explicitly parametrise the stellar component with an $M/L$ in physical units. Instead, we scale the normalised Gaussian components by a free parameter $\Upsilon_\kappa$ and apply the $M/L$ gradient. The amplitude of each Gaussian is then expressed as:
\begin{equation}
\Upsilon_{\kappa,i}(\sigma_i)=
\Upsilon_{\kappa}\,
\frac{A_{{\rm light},i}}{\sum_{j=1}^{N} A_{{\rm light},j}}\,
\sigma_i^{-\nabla M/L},
\end{equation}
where $\Upsilon_{\kappa,i}$ is the mass amplitude of the $i$th Gaussian.

For comparison with stellar-population expectations, we compute reference stellar masses for Chabrier and Salpeter IMFs from observed-frame F115W stellar-population mass-to-flux ratios. These reference values are only used to draw the IMF bands in Fig.~\ref{fig:composite_posterior}. We use the Flexible Stellar Population Synthesis code (FSPS; \citealt{Conroy2009FSPS,ConroyGunn2010FSPS}) at the lens redshift \(z_{\rm lens}=0.6575\) to generate single-burst stellar populations in the JWST/NIRCam F115W band, with \texttt{z\_continuous}=1, \texttt{sfh}=0, no dust extinction (\texttt{dust2}=0), no nebular emission, and stellar remnants included. We evaluate ages from \(3\) to \(7\,{\rm Gyr}\) and metallicities \(\log(Z/Z_\odot)=-0.2,0.0,0.2\), for both a Chabrier IMF and a Salpeter IMF \citep{Chabrier2003,Salpeter1955}. For each population model we convert the FSPS AB magnitude to an observed flux density in nJy and compute
\begin{equation}
\mu_{\rm IMF}=\frac{M_{\star,{\rm FSPS}}}{F_{\nu,{\rm F115W}}},
\end{equation}
where \(\mu_{\rm IMF}\) has units of \(M_\odot\,{\rm nJy}^{-1}\). The resulting ranges are \(\mu_{\rm Chabrier}=(1.66,2.73,4.02)\times10^6\,M_\odot\,{\rm nJy}^{-1}\) and \(\mu_{\rm Salpeter}=(3.08,5.07,7.49)\times10^6\,M_\odot\,{\rm nJy}^{-1}\), quoted as the minimum, median, and maximum over the age--metallicity grid. Multiplying these factors by the total F115W flux of the lens-light MGE gives the Chabrier and Salpeter reference stellar-mass bands shown in Fig.~\ref{fig:composite_posterior}.

\begin{figure*}
    \centering
    \includegraphics[width=0.85\textwidth]{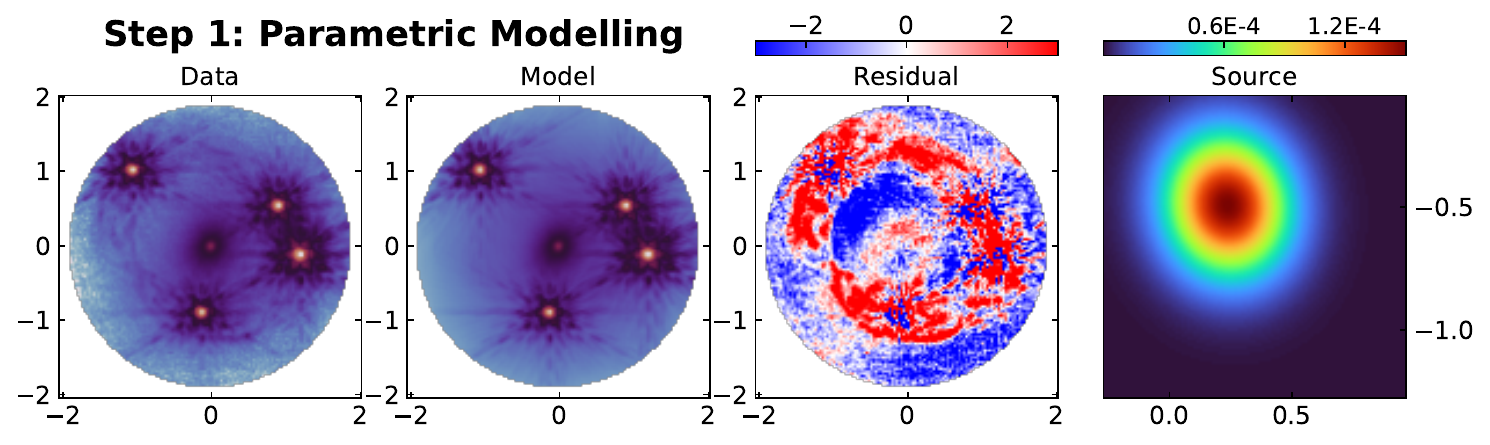}
    \includegraphics[width=0.85\textwidth]{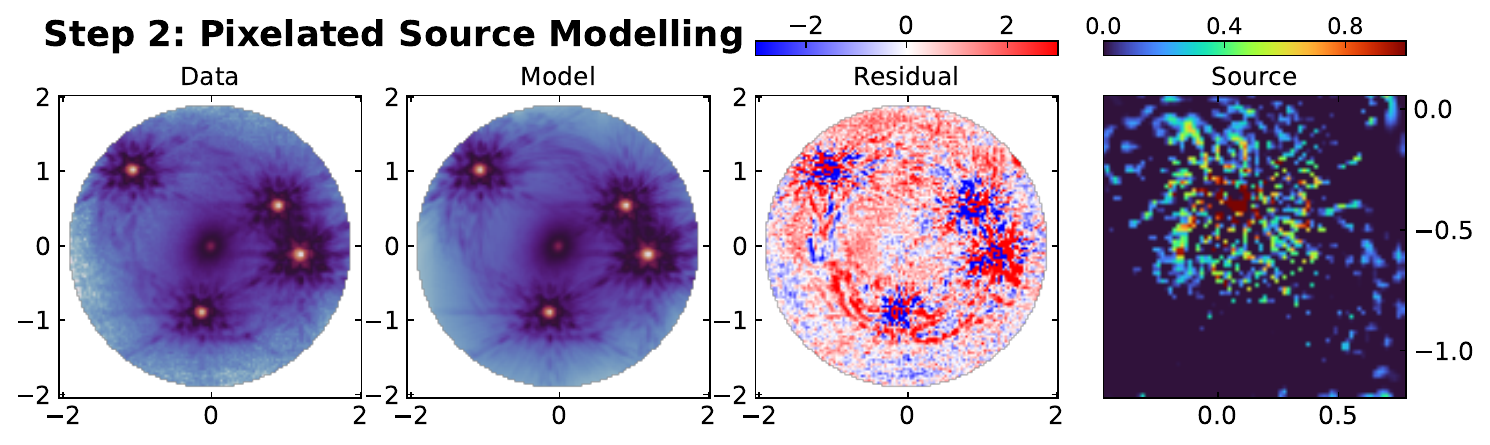}
    \includegraphics[width=0.85\textwidth]{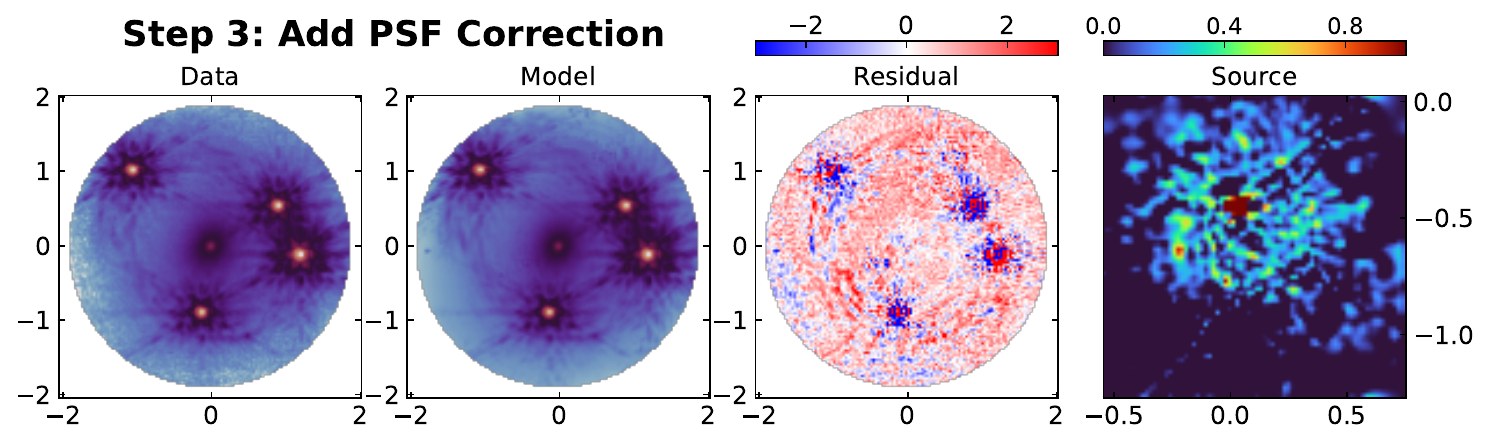}
    \includegraphics[width=0.85\textwidth]{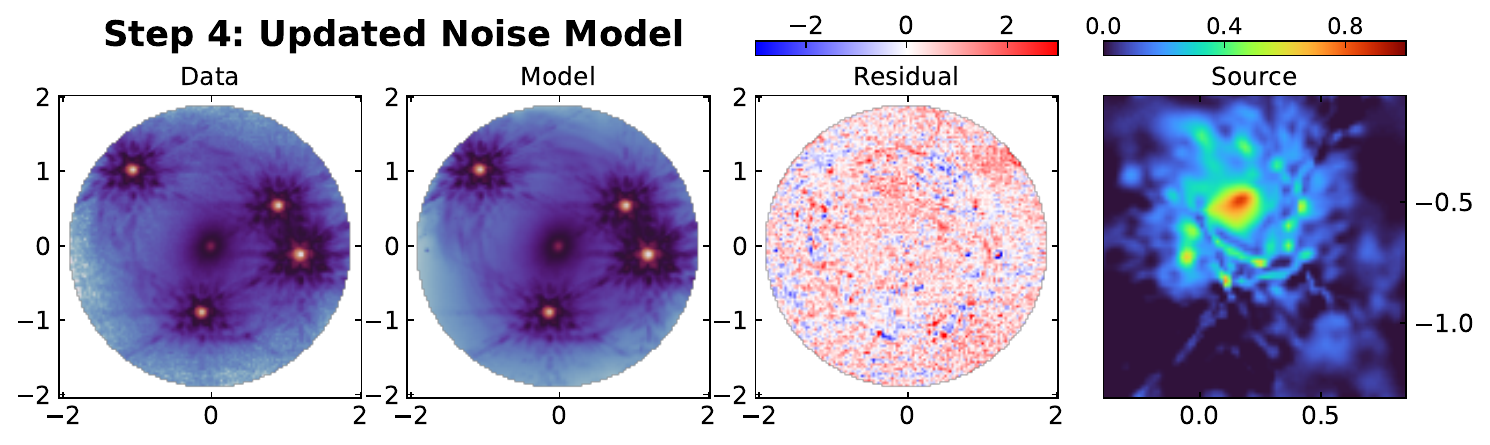}
    \caption{Stepwise image modelling of WFI2033--4723 in the JWST/NIRCam F115W band.
    Each row corresponds to one modelling stage, and the columns show, from left to right, the lens-light-subtracted data, the lensed image model, the normalized residuals, and the reconstructed source. 
    Step 1 uses a parametric source-light model.
    Step 2 replaces the source with a pixelated source reconstruction.
    Step 3 adds a correction to the PSF kernel.
    Step 4 uses the PSF-corrected model with an updated noise map that includes additional PSF-related uncertainty. All axes are given in arcseconds.
    }
    \label{fig:initialisation_sequence}
    \vspace{-0.5cm}
\end{figure*}

\subsubsection{Dark matter mass component}
The three-dimensional radial profile of the dark matter is described as an elliptical gNFW profile parametrised as:
\begin{equation}
\rho(R)=\rho_s\left(\frac{R}{r_s}\right)^{-\gamma_{\rm in}}
\left(1+\frac{R}{r_s}\right)^{\gamma_{\rm in}-3},
\end{equation}
where $\rho_s$ is the characteristic density at the scale radius $r_s$, and
$\gamma_{\rm in}$ is the inner density slope (cuspy if $\gamma_{\rm in}>1$; cored if $\gamma_{\rm in}<1$).
For $\gamma_{\rm in}=1$, the profile reduces to the classical NFW.

In \textsc{jax-lensing-profiles}, \(\kappa_{s,\rm halo}\) follows the convention of \citet{Keeton2001} and is defined as
\begin{equation}
    \kappa_{s,\rm halo}=\frac{\rho_s R_s}{\Sigma_{\rm crit}}.
\end{equation}
The gNFW halo is therefore parametrized by \(\kappa_{s,\rm halo}\), \(R_s\), and the inner slope \(\gamma_{\rm in}\).

The gNFW profile is approximated using a 3D MGE with 20 components, i.e. the dark matter mass distribution is represented as a sum of concentric elliptical Gaussian basis functions \citep{Emsellem1994_MGE, Cappellari2002_MGE, Shajib2019MGE}. A full description of the 3D MGE implementation is provided in \cite{Tian2026}.

\subsection{Initialisation}

Lens modelling of these data is a high-dimensional inference problem. As introduced in the previous sections, in the pixelated stages our model includes the lens mass, the lens light, the point sources, the source power-spectrum hyperparameters, the source pixels, and the PSF-correction field. Directly initialising the final HMC run in this full parameter space is inefficient, so we first use SVI to build a sequence of increasingly flexible models and to provide a stable starting point for the final sampling.

{\tt NumPyro} implementation of SVI \citep[see][]{WingateWeber2013} is built on JAX\footnote{\url{https://github.com/google/jax}} for efficient automatic differentiation. SVI approximates the posterior with a tractable variational distribution by optimising the evidence lower bound (ELBO). The variational guide is not intended to represent the final posterior in this high-dimensional problem, but it is much cheaper than MCMC and provides robust initial values for the subsequent HMC analysis.

Optimisation is performed with the AdaBelief optimiser \citep{Zhuang2020} and a two-stage exponential learning-rate schedule. The learning rate is initialised to \(10^{-2}\) and decays exponentially with decay rate 0.99 and transition steps 200 for the first half of the optimisation. At the midpoint we switch to a second exponential decay with the same decay rate and transition steps 10, initialised to the learning rate at the boundary to ensure continuity. The parametric stage uses a low-rank multivariate normal guide, while the pixelated stages use a diagonal normal guide initialised from the previous SVI median with an initial scale of 0.01.

Because the arcs are blended with both the lens light and the quasar images, we model the lens light, point sources, and lensed host-galaxy arcs simultaneously throughout the initialisation. The lens light is described by five elliptical Gaussian luminosity profiles. In the initial SVI stages, the inner three Gaussian components are fixed to the lens-light solution obtained from a previous corrected-PSF light fit, while the outer two Gaussian components are allowed to vary. 

We run eight independent SVI chains in each optimisation stage. The initialisation proceeds in four stages before the final HMC sampling. First, we fit a fully parametric model in which the lens light is described by five elliptical Gaussian components, the source is described by an elliptical Gaussian profile, and the four quasar images are modelled as point sources. This provides an initial solution for the lens mass, lens light, point source positions and fluxes, and the smooth source morphology.

Second, we replace the parametric source with a pixelated source model while keeping the PSF fixed. The pixel grid is constructed from the source-plane size inferred from the parametric solution, and the lens mass, lens light, and point source parameters are initialised from the first stage. Third, we activate the PSF correction and refit the pixelated model, initialising the lens mass, lens light, point sources, source pixels, and source power-spectrum hyperparameters from the fixed-PSF pixelated solution. In the end, we repeat the pixelated SVI using the corrected PSF together with the updated RMS map that includes the extra PSF-noise term described above. This final SVI stage gives the starting point used for the HMC sampling.

The role of the PSF stages is illustrated in Fig.~\ref{fig:initialisation_sequence}. With a fixed PSF, the pixelated source reconstruction contains substantial high-frequency structure, indicating that part of the source model is compensating for PSF mismatch around the bright quasar images. Applying the PSF correction already moves the reconstruction toward a smoother and more physically plausible face-on disk-like morphology. After adding the extra PSF-noise map and refitting with the corrected PSF, residual PSF mismatch no longer dominates the likelihood in the highest-surface-brightness pixels, allowing the extended arc information to drive the source reconstruction more directly.

\subsection{Bayesian inference}

\begin{figure*}
    \centering
    \includegraphics[width=1\textwidth]{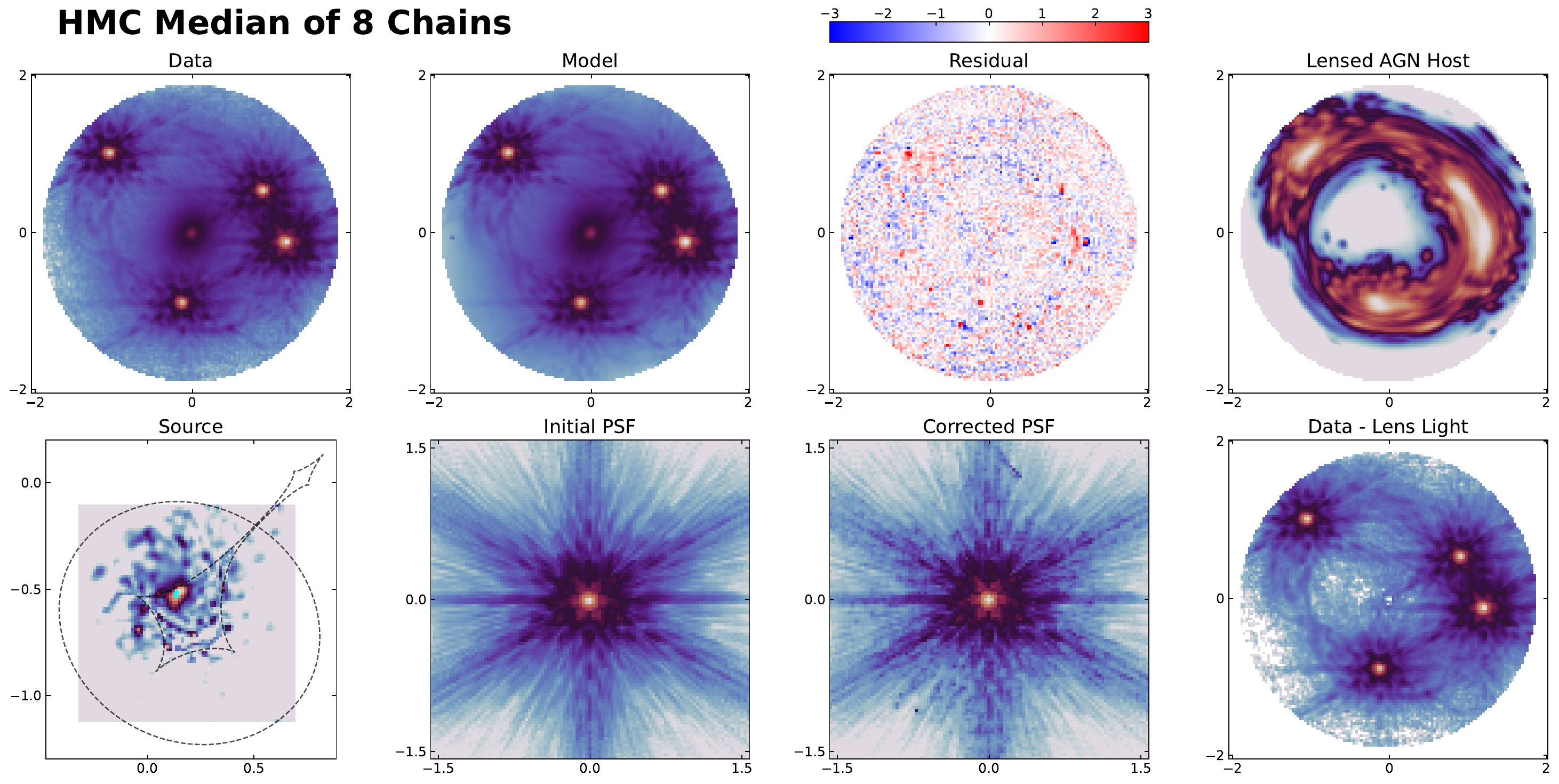}
    \vspace{-0.4cm}
    \caption{EPL baseline HMC lens-model result for WFI2033--4723. The figure shows the posterior-median image-plane reconstruction obtained from the deterministic quantities saved during sampling. The panels summarize the fit to the JWST/NIRCam F115W image, including the image-plane model, normalized residuals, reconstructed source, and the PSF-related quantities used in the HMC analysis. If instead an image-plane model is generated from the median value of each source pixel, the residual map becomes globally positive, similar to the last panel of Fig.~\ref{fig:initialisation_sequence}; this happens because the positivity constraint makes the source-pixel posterior non-Gaussian, so the collection of marginal median source-pixel values does not correspond to the posterior-median model. Without imposing source positivity, the SVI-stage residuals will not show this global offset. All axes are given in arcseconds.}
    \label{fig:hmc_median_model}
    \vspace{-0.5cm}
\end{figure*}

We estimate the posterior with NumPyro's implementation of the No-U-Turn Sampler (NUTS; \citealp{HoffmanGelman2014NUTS,Phan2019NumPyro}), an HMC method \citep{1987PhLB..195..216D,brooks2011handbook} that uses gradients to propose efficient moves. Compared with traditional MCMC, HMC/NUTS typically yields low-autocorrelation chains, needs fewer warm-up steps, and scales well to high-dimensional posteriors. To improve robustness in the high-dimensional pixelated model, we combine NUTS with an HMC-within-Gibbs scheme \citep{Krawczyk2024MultiHMCGibbs}: blocks of parameters are updated in turn while the remaining parameters are held fixed, following a Gibbs-like schedule \citep[cf.][]{GelmanBDA3}. For all production runs we inspect the rank-normalised \(\hat r\), effective sample sizes, and the fraction of divergent transitions to assess convergence.

\subsubsection{EPL baseline HMC}

The EPL baseline HMC is initialised from the final SVI solution described above. The main purpose of the HMC-within-Gibbs sampler in this stage is to separate the lens modelling parameters from the pixelized PSF correction. In practice, each Gibbs cycle alternates between updating the lens mass, light, source, and point source parameters conditioned on the current PSF, and updating the PSF-correction field conditioned on the current lens model. This makes the PSF correction an on-the-fly part of the posterior inference. The dense mass-matrix blocks used for the EPL and composite HMC runs are listed in Table~\ref{tab:hmc_blocks}. For the EPL baseline, we run eight vectorised chains, each with 2000 warm-up steps followed by eight batches of 1000 posterior samples.

\begin{table}
\centering
\small
\renewcommand{\arraystretch}{1.1}
\setlength{\tabcolsep}{4pt}
\begin{tabular}{@{}p{0.35\columnwidth}p{0.57\columnwidth}@{}}
\hline
\textbf{Block} & \textbf{Parameters in dense block} \\
\hline
\multicolumn{2}{l}{\emph{EPL baseline}}\\
Source hyperparameters & \(n_{\rm src}, \rho_{\rm src}, \sigma_{\rm src}\) \\
Lens light & Gaussian amplitudes, widths, and ellipticities \\
Point sources & quasar image positions and fluxes \\
Mass model & EPL profile, external shear, and free SIS perturbers \\
PSF correction & pixelized PSF-correction field; separate Gibbs block \\
\hline
\multicolumn{2}{l}{\emph{Composite model}}\\
Source hyperparameters & \(n_{\rm src}, \rho_{\rm src}, \sigma_{\rm src}\) \\
Mass and cosmology & stellar \(M/L\), \(M/L\) gradient, gNFW halo, shear, SIS perturbers, \((\Omega_{\rm m},H_0)\), and \(\kappa_{\rm ext}\) \\
Point sources & quasar image positions and fluxes \\
\hline
\end{tabular}
\caption{Dense mass-matrix blocks and the separate PSF-correction Gibbs block used in the HMC sampling.}
\label{tab:hmc_blocks}
\end{table}

\begin{figure*}
    \centering
    \includegraphics[width=0.8\textwidth]{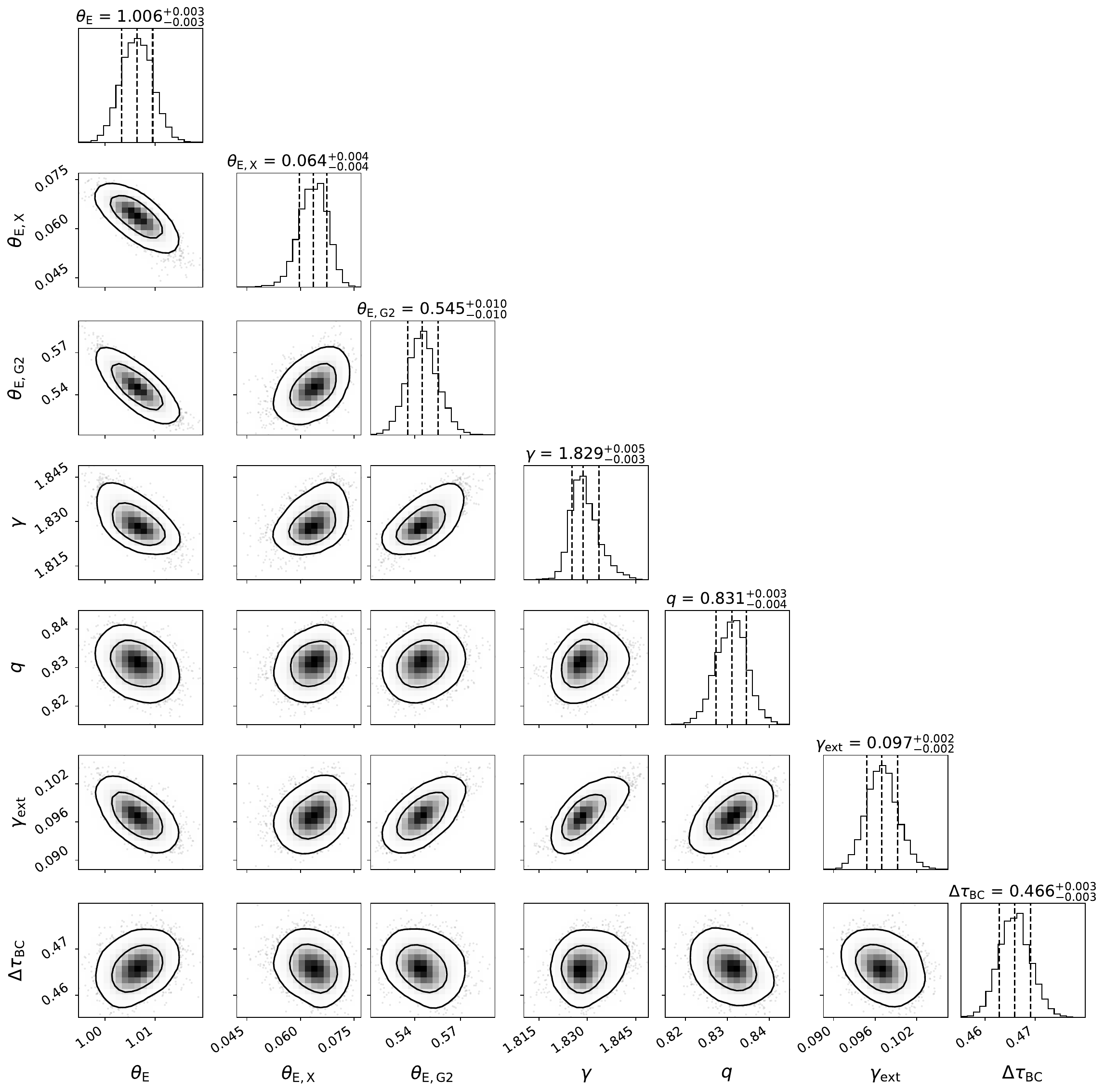}
    \begingroup
    \large
    \renewcommand{\arraystretch}{1.25}
    \setlength{\tabcolsep}{7pt}
    \begin{tabular}{@{}ccccccccc@{}}
    \hline
    $\theta_{\rm E}$ & $\gamma$ & $q$ & $\gamma_{\rm ext}$ & $\theta_{\rm E,X}$ & $\theta_{\rm E,G2}$ & $\Delta\tau_{\rm BC}$ & $\Delta\tau_{\rm BA1}$ & $\Delta\tau_{\rm BA2}$ \\
    \hline
    $1.006^{+0.003}_{-0.003}$ & $1.829^{+0.005}_{-0.003}$ & $0.831^{+0.003}_{-0.004}$ & $0.097^{+0.002}_{-0.002}$ & $0.064^{+0.004}_{-0.004}$ & $0.545^{+0.010}_{-0.010}$ & $0.466^{+0.003}_{-0.003}$ & $0.305^{+0.004}_{-0.004}$ & $0.325^{+0.005}_{-0.005}$ \\
    \hline
    \end{tabular}
    \endgroup
    \caption{Posterior distributions of the HMC chain. The corner plot shows the 2D posterior distributions for the EPL mass model. The table reports posterior medians and 16th--84th percentile uncertainties, including the three fpd used for the time-delay analysis.}
    \label{fig:epl_posterior}
    \vspace{-0.5cm}
\end{figure*}

\begin{figure*}
    \centering
    \includegraphics[width=1\textwidth]{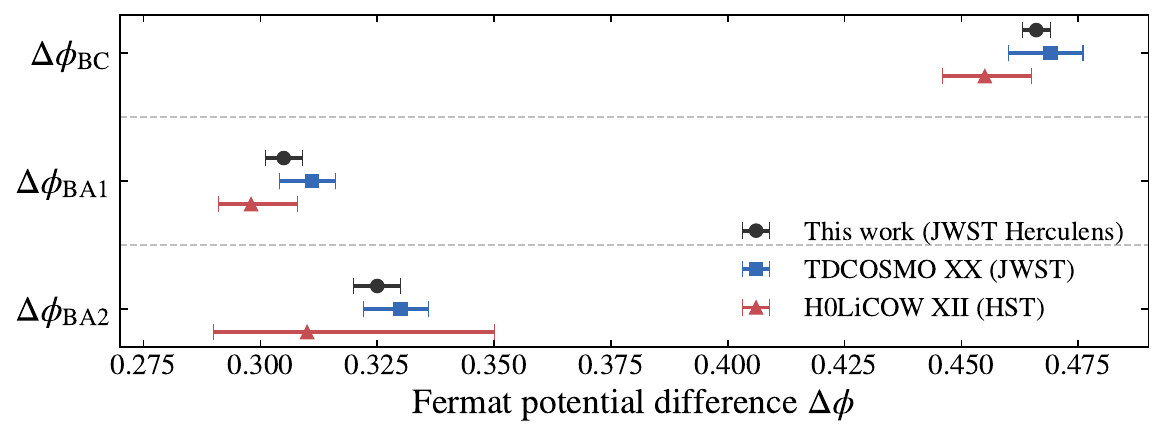}
    \vspace{-0.5cm}
    \caption{Comparison of fpd for the three image-pair combinations. The error bars show the median and 16th--84th percentile. Our result is compared with the JWST analysis from TDCOSMO XX and the HST-based H0LiCOW XII result.}
    \label{fig:fermat_comparison}
    \vspace{-0.5cm}
\end{figure*}

\subsubsection{Composite HMC}

The star+gNFW inference is initialised chain-by-chain from the EPL HMC result. For each EPL chain, we construct a preliminary stellar-plus-NFW model whose convergence approximately matches the corresponding EPL convergence over the lensed arc on the image plane. This matching is done with a least-squares fit to the convergence residuals in the image plane, varying the stellar normalisation, halo normalisation, and halo ellipticity to find the composite profile closest to the EPL convergence map. During this step the initial halo has \(R_s=5^{\prime\prime}\), \(\gamma_{\rm in}=1\), and the same centre as the corresponding EPL chain. This provides a one-to-one mapping between each EPL chain and an initial composite mass model. The remaining variables are transferred from the corresponding EPL chain where possible.

We then run a sequence of star+gNFW SVI optimisations before the final HMC:
\begin{itemize}
\item The first composite SVI keeps \(R_s=5^{\prime\prime}\), fixes the halo centre and shear to the EPL values, and uses a constant stellar \(M/L\).
\item The second stage releases \(R_s\), the halo centre and shear, the main SIS perturbers, and the quasar point source parameters.
\item The final stage enables the radial stellar \(M/L\) gradient and the time-delay likelihood under a specified cosmological prior.
\end{itemize}

The final composite HMC is then initialised from this SVI solution and sampled with the composite dense blocks listed in Table~\ref{tab:hmc_blocks}. We repeat the composite inference separately for the Pantheon+SH0ES and Planck+DESI cosmological priors. Following \citet{Tian2026}, we adopt a broad gNFW scale-radius prior \(R_s\sim\mathcal{U}(2,20)\) arcsec, which avoids imposing a direct concentration--mass prior while keeping the halo scale radius within a physically motivated range.

\subsection{EPL baseline and Fermat-potential constraints}
\label{sec:epl_fermat_results}

Figure~\ref{fig:hmc_median_model} shows the EPL baseline lens-model result, obtained by taking the median of the deterministic lens-model quantities saved during sampling. We successfully modelled the quasar images and the extended host-galaxy arcs to the noise level. The reconstructed host is a nearly face-on disk galaxy, with a bright central component and low-surface-brightness spiral-arm-like structure visible in the source plane. Several compact knots are also recovered around the disk.

The sampled PSF correction mainly modifies the inner PSF core. Relative to the modelled initial PSF obtained from the field-star SVI fit, the corrected PSF has a lower peak value and a larger FWHM. The corrected PSF also contains low-level noise and pseudo-structures, which likely reflect the limited signal-to-noise ratio of the four quasar point source images for constraining a pixelized PSF correction. In particular, the diagonal feature above the PSF core is likely associated with residuals from imperfect modelling in the overlap region between the lower-right side of image~A2 and the region above image~A1. Because this structure carries little flux, we do not expect it to have a significant impact on the lens modelling. In future work, a simple mask at this location during the PSF-correction step would remove this feature.

Figure~\ref{fig:epl_posterior} shows the posterior distribution of the EPL baseline model, and the embedded table summarises the marginalised constraints on the main lens parameters and fpd. The eight independent HMC chains are concatenated directly when producing this posterior summary. The \(\hat r\) values of all parameters are below 1.05, even though the eight chains are initialised from different starting points and each chain samples its own pixelized PSF-correction field. The chains still converge to the same posterior distribution, showing that the on-the-fly PSF correction does not lead to chain-dependent lens solutions.

Figure~\ref{fig:fermat_comparison} compares the fpd from the EPL baseline with the HST-based H0LiCOW XII analysis and the JWST-based TDCOSMO XX analysis. For the three image-pair combinations entering the time-delay likelihood, we obtain \(\Delta\phi_{\rm BC}=0.466^{+0.003}_{-0.003}\), \(\Delta\phi_{\rm BA1}=0.305^{+0.004}_{-0.004}\), and \(\Delta\phi_{\rm BA2}=0.325^{+0.005}_{-0.005}\). The comparison shows that our EPL baseline is consistent with both previous analyses within the quoted uncertainties. This agreement is important for the composite analysis below: it shows that the EPL baseline recovers the established Fermat-potential scale for WFI2033--4723 before we replace the total power-law profile with the more flexible stellar-plus-dark-matter decomposition.

\subsection{Composite model}

\begin{figure*}
    \centering
    \includegraphics[width=1\textwidth]{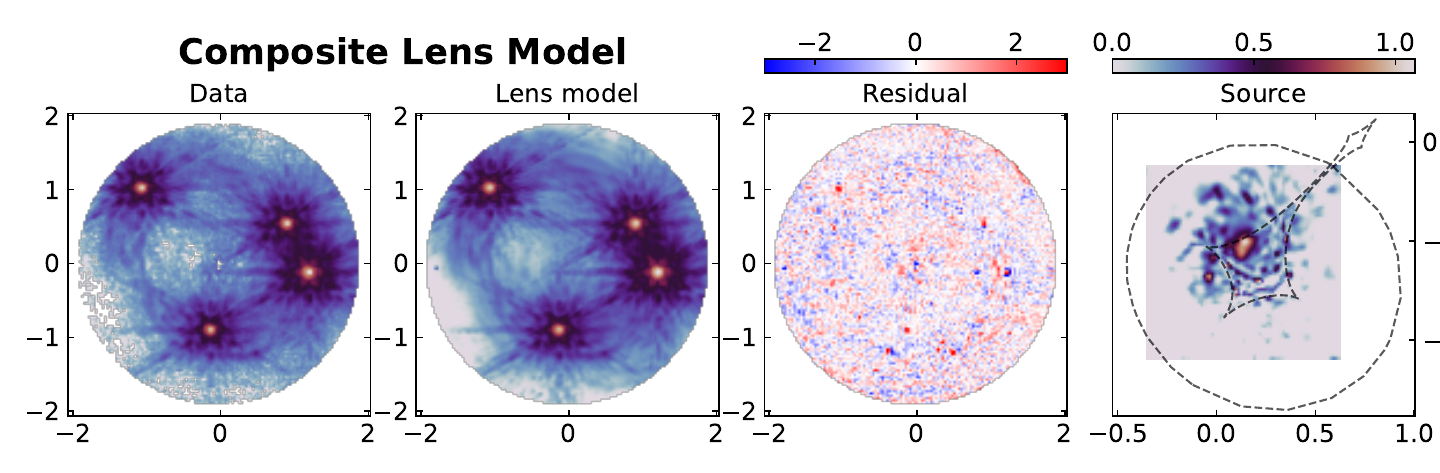}
    \vspace{-0.4cm}
    \caption{Composite star+gNFW lens-model result. The four panels show the posterior-median lens model. From left to right, the panels show the observed image, model image, normalized residual, and reconstructed source on the source plane. All axes are given in arcseconds.}
    \label{fig:composite_lens_model}
    \vspace{-0.5cm}
\end{figure*}

\begin{figure*}
    \centering
    \includegraphics[width=0.75\textwidth]{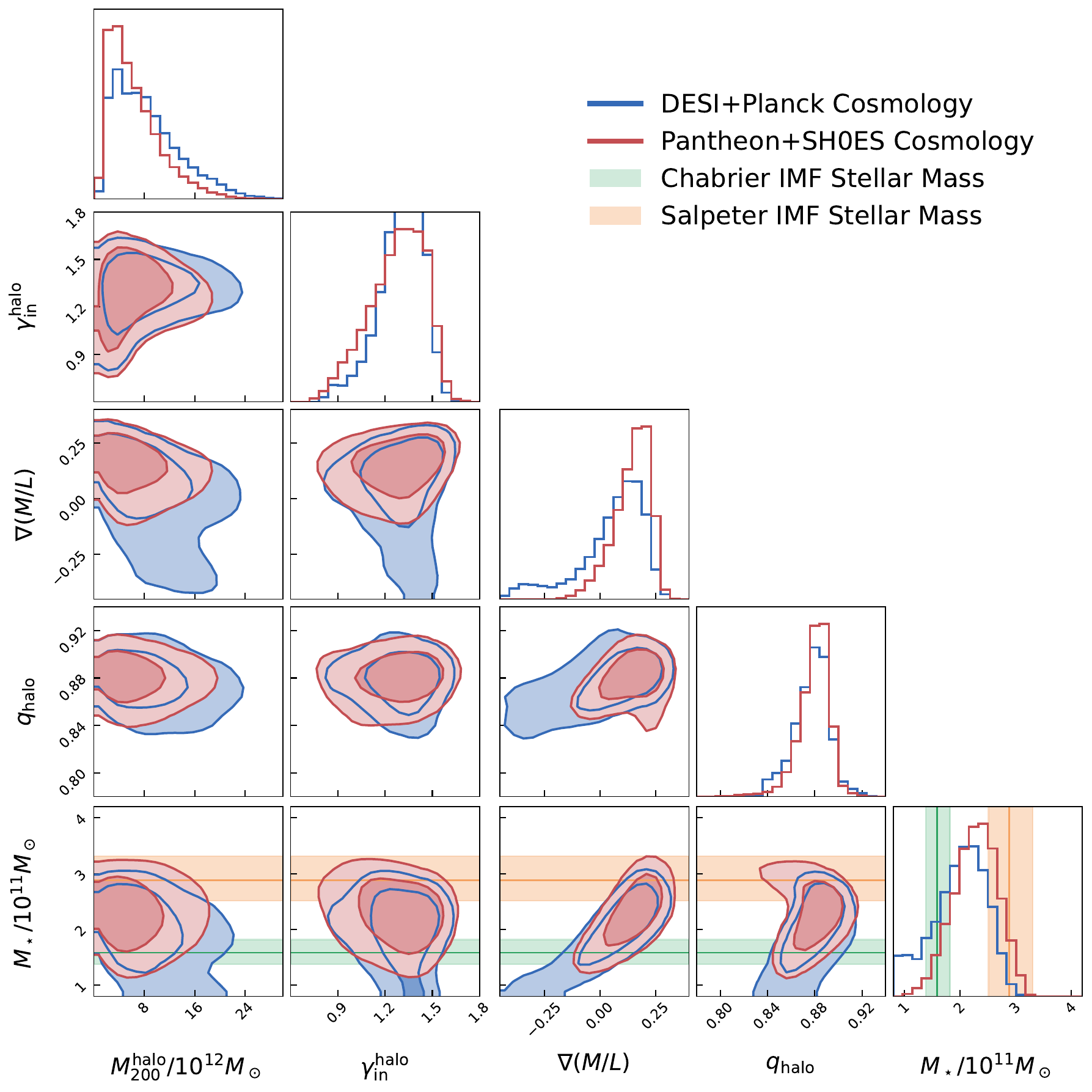}
    \caption{Selected posterior distributions for the composite star+gNFW time-delay model under the two external cosmological priors. Diagonal panels show the one-dimensional marginalised posteriors, while off-diagonal panels show the two-dimensional posteriors with credible contours. The displayed parameters include the stellar mass, radial stellar \(M/L\) gradient, and the main gNFW halo quantities used for the physical interpretation. The shaded bands indicate the Chabrier and Salpeter IMF expectations.}
    \label{fig:composite_posterior}
    \vspace{-0.5cm}
\end{figure*}

\begin{table*}
\centering
\caption{Posterior constraints for the composite star+gNFW time-delay models under the two external cosmological priors. Reported values are the marginal posterior medians, with uncertainties corresponding to the central 68\% credible intervals.}
\label{tab:composite_posterior}
{\renewcommand{\arraystretch}{1.3}
\begin{tabular}{lrrrrrrrrr}
\hline
Model & $M_{200}^{\rm halo}/10^{12}M_\odot$ & $R_s^{\rm halo}$ & $\gamma_{\rm in}^{\rm halo}$ & $q_{\rm halo}$ & $\nabla (M/L)$ & $M_\star/10^{11}M_\odot$ & $\theta_{\rm E}^{\rm X}$ & $\gamma_{\rm ext}$ & $\kappa_{\rm ext}$ \\
\hline
Planck+DESI & $7.52^{+6.23}_{-4.14}$ & $9.21^{+6.75}_{-5.10}$ & $1.32^{+0.12}_{-0.15}$ & $0.88^{+0.01}_{-0.02}$ & $0.09^{+0.10}_{-0.17}$ & $2.03^{+0.43}_{-0.62}$ & $0.057^{+0.002}_{-0.002}$ & $0.104^{+0.005}_{-0.007}$ & $0.139^{+0.059}_{-0.071}$ \\
Pantheon+SH0ES & $5.45^{+4.94}_{-2.80}$ & $8.32^{+7.96}_{-4.93}$ & $1.30^{+0.16}_{-0.21}$ & $0.88^{+0.01}_{-0.01}$ & $0.16^{+0.06}_{-0.10}$ & $2.30^{+0.38}_{-0.43}$ & $0.060^{+0.002}_{-0.002}$ & $0.114^{+0.005}_{-0.006}$ & $0.121^{+0.052}_{-0.057}$ \\
\hline
\end{tabular}}
\end{table*}

\begin{figure*}
    \centering
    \includegraphics[width=0.7\textwidth]{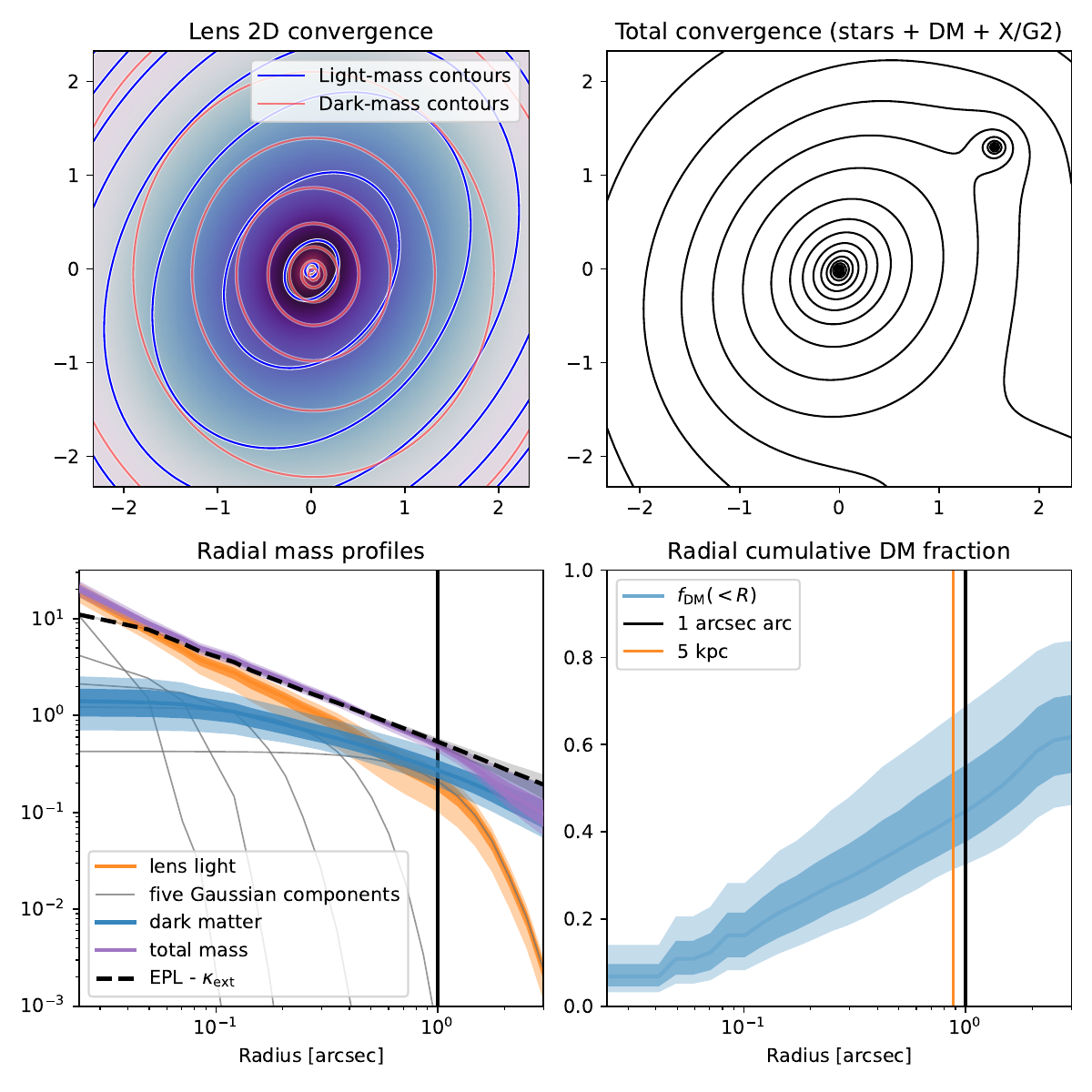}
    \vspace{-0.4cm}
    \caption{The stellar and dark-matter mass decomposition. \textit{Top-left:} two-dimensional convergence map of the median composite lens model, with the total convergence shown by the colour scale and the stellar and dark-matter components overlaid as contours. \textit{Top-right:} contours of the total convergence including the nearby perturbers. \textit{Bottom-left:} radial convergence profiles of the stellar component, dark-matter halo, and total composite mass profile. The grey curves show the contribution of each stellar-light Gaussian component, the solid coloured curves show the posterior median profiles, the shaded bands indicate the central 68\% credible regions, and the black dashed curve shows the EPL baseline profile after subtracting the external convergence, \(\kappa_{\rm EPL}-\kappa_{\rm ext}\). \textit{Bottom-right:} cumulative dark-matter fraction \(f_{\rm DM}(<R)\) as a function of radius.}
    \label{fig:composite_lens_model_convergence}
    \vspace{-0.5cm}
\end{figure*}

\label{sec:composite_results}

Figure~\ref{fig:composite_lens_model} shows the posterior-median result for the star+gNFW time-delay model. The composite model fits the quasar images and extended arcs at a level comparable to the EPL baseline. The lower panels show the corresponding radial mass profiles: the stellar and halo terms can trade off against each other while preserving a similar total convergence near the Einstein ring, which is the MST-like freedom that the time-delay information is intended to constrain. The quantitative constraints are summarised in Table~\ref{tab:composite_posterior}. Figure~\ref{fig:composite_posterior} shows the selected posterior distributions of the main stellar and halo parameters under the two cosmological priors; the diagonal panels show the one-dimensional marginalised posteriors, the off-diagonal panels show the joint constraints, and the IMF bands provide a visual reference for interpreting the stellar normalisation.

\subsubsection{Stellar component}

The stellar component should be the part of the decomposition affected by the adopted \(H_0\) prior, since \(H_0\) is linearly related to the MST rescaling. The inferred stellar mass increases from \(M_\star=2.03^{+0.43}_{-0.62}\times10^{11}M_\odot\) for Planck+DESI to \(M_\star=2.30^{+0.38}_{-0.43}\times10^{11}M_\odot\) for Pantheon+SH0ES, and the radial \(M/L\) gradient shifts from \(\nabla(M/L)=0.09^{+0.10}_{-0.17}\) to \(0.16^{+0.06}_{-0.10}\). In both cases the stellar normalisation lies between the Chabrier and Salpeter IMF bands shown in Fig.~\ref{fig:composite_posterior}. This is lighter than the Salpeter-like stellar normalisation inferred for the Jackpot lens by \citet{Tian2026}, but still heavier than a purely Chabrier expectation. A similarly intermediate, heavier-than-Chabrier but sub-Salpeter IMF normalisation has also been reported for the central region of the Euclid lens NGC~6505 \citep{Euclid2025NGC6505}. We find a mild preference for a positive \(M/L\) gradient. In this work we allow \(\nabla(M/L)\in[-0.6,0.6]\), whereas \citet{Tian2026} restricted the gradient to the negative range \([-0.6,0]\). If the same negative-gradient prior were imposed here, the solution would be pushed toward the zero-gradient boundary and would be consistent with the approximately flat \(M/L\) profile found for the Jackpot. The posterior also shows a strong degeneracy between the stellar mass normalisation and the \(M/L\) gradient; solutions closer to the Chabrier stellar mass band are associated with gradients closer to zero. We interpret the data as favouring a moderately heavy stellar population, with a slight hint of a strong radial IMF or stellar-population gradient (but not definitive).

\begin{figure*}
    \centering
    \includegraphics[width=1\textwidth]{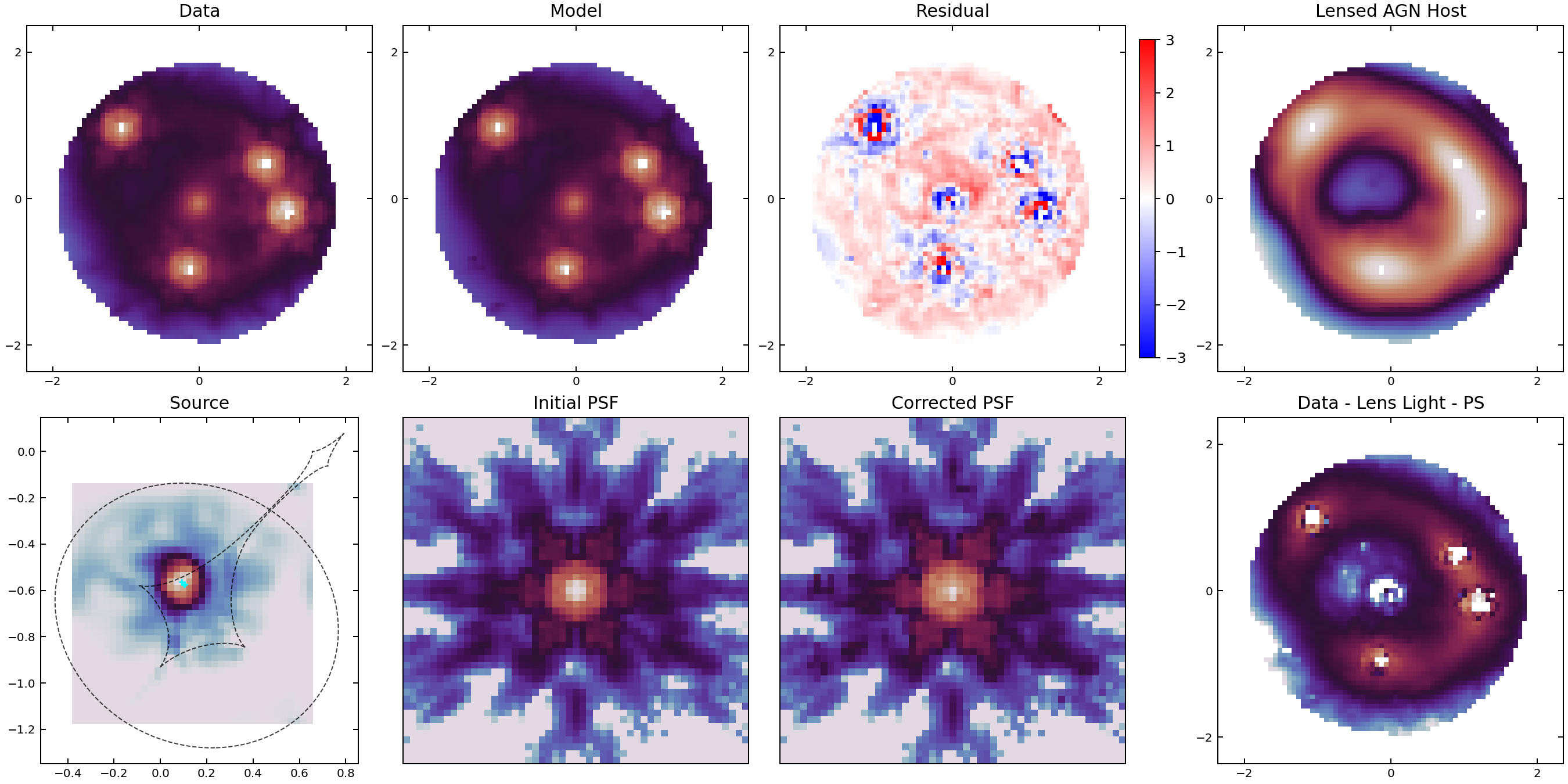}
    \vspace{-0.4cm}
    \caption{Fixed-mass F356W light-model result for WFI2033--4723. The panels show the JWST/NIRCam F356W data, best-fitting image-plane model, residuals in units of the noise map, lensed host-galaxy light without the quasar point source components, reconstructed source with caustics, the modelled initial F356W PSF from the field-star SVI fit, the PSF after correction, and the image after subtracting both the lens light and quasar PSF components. The mass model is fixed to the F115W EPL HMC result projected into the F356W cutout frame, while NaN/saturated pixels are masked in the fit. All axes are given in arcseconds.}
    \label{fig:f356w_fixedmass_model}
    \vspace{-0.5cm}
\end{figure*}

\begin{figure}
    \centering
    \includegraphics[width=0.5\textwidth]{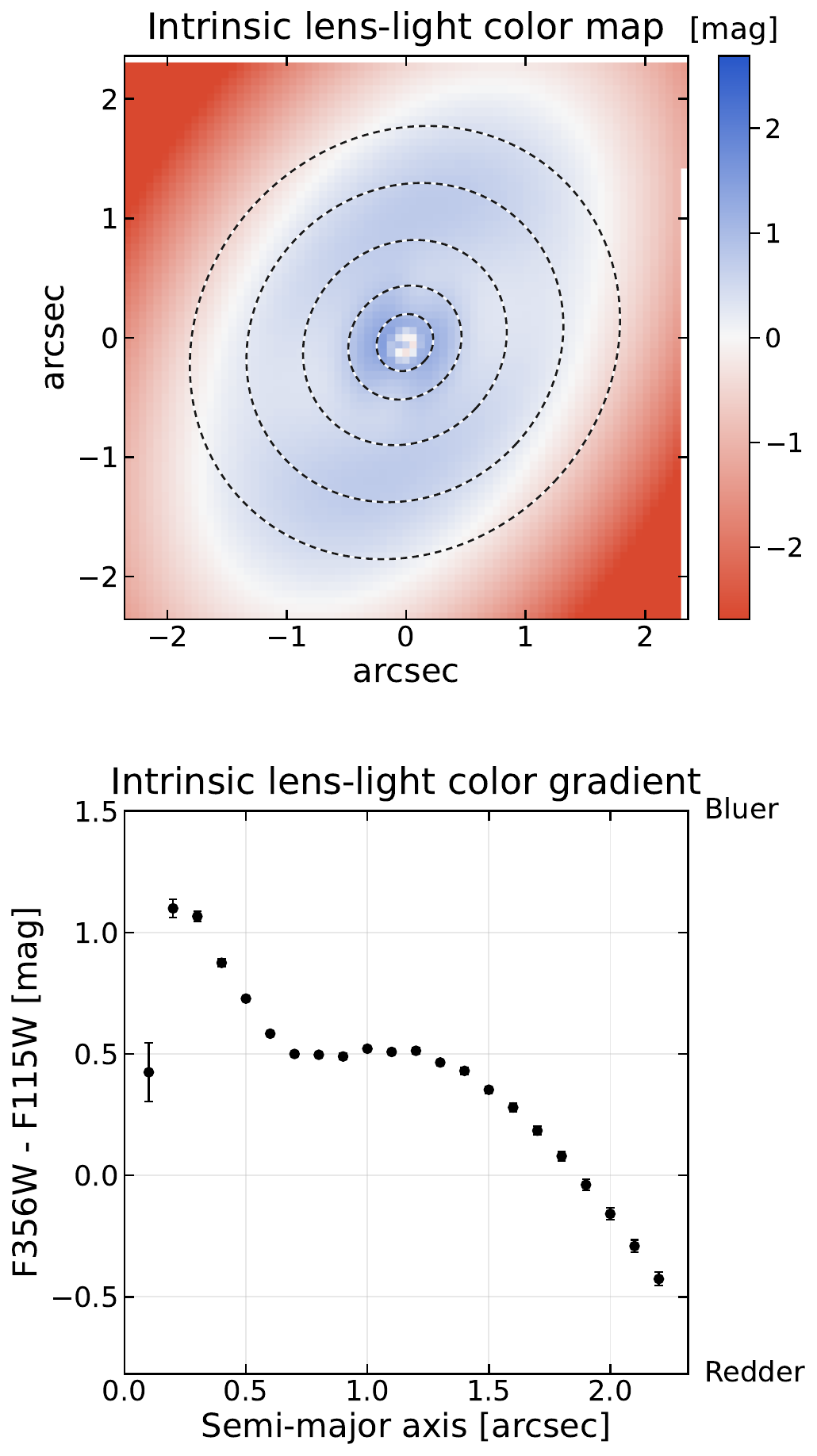}
    \vspace{-0.4cm}
    \caption{Intrinsic lens-light color gradient of WFI2033. The upper panel shows the two-dimensional color map measured from the intrinsic, pre-PSF-convolution lens-light models in F356W and F115W, aligned on the F356W grid. Black dashed ellipses mark the apertures used for the elliptical photometry. The color is defined as $F356W-F115W=-2.5\log_{10}(I_{\rm F356W}/I_{\rm F115W})$; therefore orange-red, negative values correspond to relatively stronger F356W emission and a yellower/redder color, while blue, positive values correspond to a bluer color. The lower panel shows the corresponding elliptical-aperture color profile as a function of semi-major axis. The outward decrease in $F356W-F115W$ indicates that the lens light becomes progressively yellower/redder at larger radii.}
    \label{fig:lenslight_color_gradient}
    \vspace{-0.5cm}
\end{figure}

As a qualitative check on this interpretation, we also perform a simple PSF-corrected SVI model on an F356W cutout. In this test, we fix all mass parameters to the posterior median of the F115W EPL HMC model and fit only the lens light, source light, point sources, and PSF correction. Because the F356W quasar images are saturated, the central PSF correction is not fully reliable; this likely causes the over-subtraction of the lens light in the core and the negative central residuals seen in Fig.~\ref{fig:f356w_fixedmass_model}. These central residuals do not affect our qualitative use of the model. We then measure the intrinsic \(F356W-F115W\) colour profile from the F356W and F115W lens-light models, as shown in Fig.~\ref{fig:lenslight_color_gradient}. The deflector becomes redder at larger radius, implying that the outer stellar population could have a higher \(M/L\). Although we do not perform a full stellar-population analysis, this outward reddening is consistent with the positive \(M/L\) gradient inferred by the composite lens model.

\subsubsection{Dark matter halo}

\begin{figure*}
    \centering
    \includegraphics[width=0.8\linewidth]{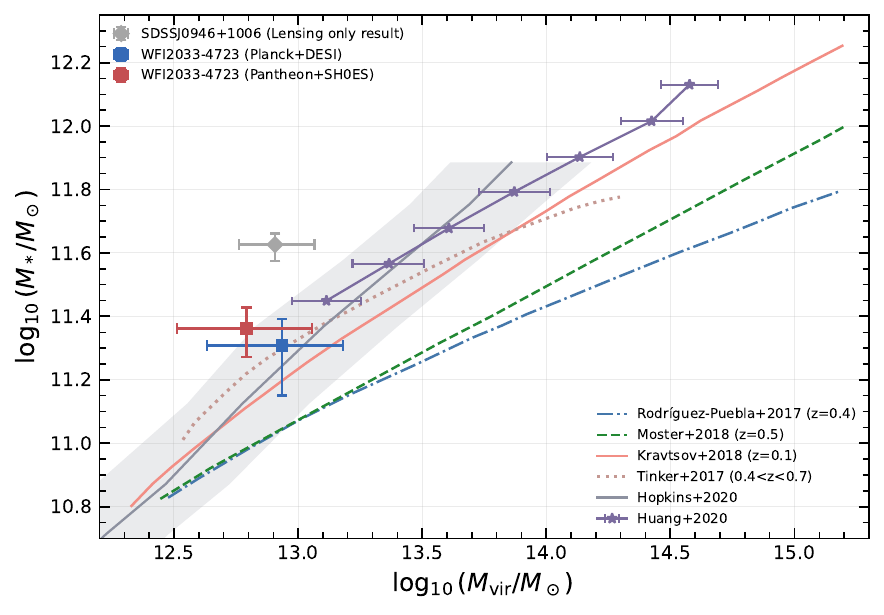}
    \caption{Stellar-to-halo mass relation for massive lens galaxies. The grey band and grey line show the relation from \citet{Hopkins2020}, and the other coloured curves show literature relations from \citet{RodriguezPuebla2017}, \citet{Moster2018}, \citet{Kravtsov2018}, and \citet{Tinker2017}. The purple points show the massive-galaxy measurements from \citet{Huang2020}. We overplot the Jackpot constraints from \citet{Tian2026} for the lensing-only model and for the model including kinematics and the concentration--mass prior. The WFI2033--4723 constraints from this work are shown for the Planck+DESI and Pantheon+SH0ES cosmological priors. Halo masses are converted from \(M_{200}\) to \(M_{\rm vir}\) for comparison with the literature relations.}
    \label{fig:shmr}
    \vspace{-0.5cm}
\end{figure*}

The dark-matter halo is more stable with respect to the adopted cosmological prior. We infer $\gamma_{\rm in}=1.32^{+0.12}_{-0.15}$ for Planck+DESI and $\gamma_{\rm in}=1.30^{+0.16}_{-0.21}$ for Pantheon+SH0ES, while the projected halo axis ratio remains $q_{\rm halo}\simeq0.88$ in both runs. This inner slope is noticeably steeper than a standard NFW cusp and also steeper than the Jackpot result of \citet{Tian2026}, where the same star+gNFW framework gave an NFW-like halo with $\gamma_{\rm in}\simeq1.04$. This comparison should be interpreted together with the posterior degeneracies shown in Fig.~\ref{fig:composite_imf_corner}: $\gamma_{\rm in}$ is strongly correlated with both the stellar mass normalisation and the halo scale radius. In particular, $R_s$ is not tightly constrained in either cosmological run, with broad posteriors spanning much of the adopted prior range. This is also reflected in the convergence diagnostic, for which the scale radius has a slightly elevated value of $\hat{R}=1.06$, while all other model parameters have $\hat{R}<1.05$. This differs from the lensing-only Jackpot analysis of \citet{Tian2026}, where the scale radius was already constrained to be of order ten times the half-light radius. A plausible explanation is that the double-source-plane geometry of the Jackpot provides two Einstein rings, and therefore two radii at which the enclosed mass is accurately constrained, whereas WFI2033--4723 provides one main ring plus the time-delay information. The robust result here is therefore the local stellar--dark-matter decomposition and the preference for a steep inner halo slope.

\subsubsection{Stellar-to-halo mass relation}

\begin{figure}
    \centering
    \includegraphics[width=1\linewidth]{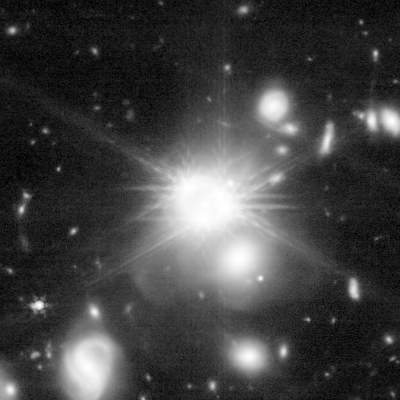}
    \caption{JWST/NIRCam F356W image of WFI2033--4723. Stretched to reveal the stellar tidal feature of the deflector galaxy. The redder band highlights the stellar light around the main lens galaxy and the nearby G2 galaxy, including diffuse shell-like tidal structures that complicate a purely smooth MGE description of the lens light.}
    \label{fig:f356w_lens_light}
\end{figure}

We use the inferred stellar and halo masses to compare WFI2033--4723 with empirical stellar-to-halo mass relations (SHMR). For comparison with the literature SHMRs in Fig.~\ref{fig:shmr}, we express the halo masses as \(M_{\rm vir}\). WFI2033--4723 lies broadly within the region occupied by massive early-type galaxies in empirical SHMR measurements and is consistent with the Jackpot constraints from \citet{Tian2026} within the current halo-mass uncertainties. The two cosmological-prior runs only mildly shift the SHMR. 

In future searches, this method could be extended to lensed AGN systems whose deflectors are disk galaxies. Recent uniform modeling of doubly imaged quasars has already provided promising examples of such systems \citep{Brady2026}. Applying this approach to disk-deflector lenses would open a new route to studying the distinctive dynamics of disk galaxies, including the role of gravitational instability in regulating their stellar mass fractions \citep{Romeo2020}.

\subsubsection{Stellar-light morphology}

Figure~\ref{fig:f356w_lens_light} shows the F356W image of WFI2033--4723. The lens galaxy and the nearby G2 galaxy are embedded in extended low-surface-brightness tidal structures, including shell-like features, indicating that the system is not a simple isolated early-type galaxy. These structures make it challenging to describe the stellar mass distribution with a small number of smooth MGE components. The MGE stellar model used in this work should therefore be viewed as an effective description of the central stellar mass traced by the lens light. A more flexible non-parametric lens-light model, or a hybrid model that treats the central galaxy and diffuse tidal shells separately, may be required to fully capture this complexity.

The wavelength used to define the stellar light is another limitation of the present analysis. The lens modelling in this work uses the high-resolution short-wavelength NIRCam image, F115W, because it provides the strongest angular-resolution constraints from the quasar images and lensed arcs; F150W would play a similar role in a short-wavelength lensing analysis. Ideally, however, the stellar mass model should be informed by the redder F356W light, which traces the old stellar population more directly and is less sensitive to stellar-population colour gradients and radial \(M/L\) variations. A future multi-band analysis should use F356W to constrain the stellar light and mass distribution, while using F115W or F150W for the high-resolution lensing constraints. However, as shown in Fig.~\ref{fig:f356w_lens_light}, the lens light is more complicated than a standard smooth MGE description, so obtaining the true distribution of the stellar mass will be challenging.

\subsection{Implication to cosmology}

Composite mass profiles have been used to constrain \(H_0\) in time-delay cosmography for more than a decade. In early H0LiCOW analyses, a stellar component tied to the observed lens light plus an NFW-like dark halo was used alongside power-law profiles to test mass-model systematics and infer time-delay distances \citep{Suyu2014,Wong2017H0LiCOWIV,Bonvin2017H0LiCOWV,Wong2020H0LiCOWXIII,Millon2020TDCOSMOI}. The broad agreement between composite and power-law models in these analyses should be interpreted with care: a fixed-\(M/L\) stellar component plus an NFW halo is still a simply parametrized profile, like an EPL, and can partially break the MSD through the imposed model assumptions (although they might not converge to the same MSD-broken solution). Later work therefore introduced stellar kinematics, including spatially resolved IFU constraints and tests of projection effects, to provide external information on the mass profile and to break or control the internal MSD \citep{Birrer2020TDCOSMOIV,Yildirim2020,Yildirim2023TDCOSMOXIII,Shajib2023TDCOSMOXII,Huang2025}. This motivates a direct test: can the more flexible star+gNFW model used in this work recover \(H_0\) without an external prior, or does it still require external information to calibrate the global mass profile?

We therefore performed an additional experiment in which \(H_0\) is inferred directly from the star-plus-dark-matter model, without imposing an external cosmological prior. This is a deliberately stringent test of the composite model: if the physical decomposition fully removed the relevant MST freedom, the time delays and imaging data should select both a plausible halo structure and a reliable time-delay distance. Instead, the result shows that the cosmology-free run does produce a converged posterior, but numerical convergence of the HMC chains is not sufficient to guarantee an unbiased cosmological inference. As shown in Fig.~\ref{fig:jampy_sigma_ap}, \(H_0\) is strongly degenerate with the dark-matter scale radius \(R_s\). In the imaging + time-delay lens model, \(R_s\) is driven toward \(2^{\prime\prime}\), close to the lower edge of the adopted prior and implausibly small for a galaxy-scale halo. This small \(R_s\) pushes the inferred \(H_0\) high. Because the present data do not independently measure the global halo scale radius, we interpret this preference for an unphysically small \(R_s\) as a manifestation of unknown modelling systematics.

For the main analysis we therefore condition the composite models on external cosmological priors and use the time delays to constrain the stellar--dark-matter decomposition. The two composite models use different cosmological assumptions but lead to similar astrophysical measurements. Both cosmological priors give stellar masses, \(M/L\) gradients, and gNFW halo parameters in the same broad range, even though the adopted \(H_0\) priors differ. 

We also compare the two cosmological-prior runs using the Bayesian information criterion,
\begin{equation}
\mathrm{BIC}=k\ln n-2\ln\hat{\mathcal{L}},
\end{equation}
where \(k\) is the number of free parameters, \(n\) is the number of data points entering the likelihood, and \(\hat{\mathcal{L}}\) is the maximum likelihood. This comparison is experimental, as we did not vary the modelling choices or random seeds to estimate an uncertainty on the BIC. The two runs have the same number of free parameters and data points, so this \(\Delta\mathrm{BIC}\) is simply the difference in the best-fit \(\chi^2\). We find \(\mathrm{BIC}_{\rm Planck+DESI}=56976.67\) and \(\mathrm{BIC}_{\rm Pantheon+SH0ES}=56927.85\), corresponding to \(\Delta\mathrm{BIC}=48.82\) in favour of Pantheon+SH0ES. Under the standard BIC interpretation, such a large value would correspond to a very strong preference within the adopted likelihood, but in the context of lens modelling this statement should be taken with care. Decomposing the score shows that the preference is dominated by the image-plane likelihood. We therefore interpret the smaller BIC of the Pantheon+SH0ES run as indicating that this prior leads to a slightly better lens-modelling solution within our adopted setup. In a high-dimensional pixel-based lens model, differences of this size can arise from small changes in the source reconstruction, lens-light subtraction, PSF treatment, or residual limitations of the mass model. Thus, even though our modelling result prefers the Pantheon+SH0ES cosmology within this particular setup, it should not be taken as a definitive result.

As an additional diagnostic, we compute the luminosity-weighted central aperture velocity dispersion for posterior samples using the axisymmetric \textsc{JAMpy} \citep{Cappellari2008,Cappellari2020}. We evaluate \(\sigma_{\rm ap}\) within a \(1^{\prime\prime}\) circular aperture. For the fixed-cosmology runs, the predicted dispersions are compact: \(\sigma_{\rm ap}=207.00^{+2.76}_{-2.93}\,{\rm km\,s^{-1}}\) for Planck+DESI and \(213.33^{+3.42}_{-3.18}\,{\rm km\,s^{-1}}\) for Pantheon+SH0ES. In contrast, the imaging-only \(H_0\)-free run gives a broader prediction, \(\sigma_{\rm ap}=223.98^{+10.33}_{-10.44}\,{\rm km\,s^{-1}}\). This is consistent with the interpretation above: once the MST-like freedom is restricted by an external cosmology, the dynamical prediction is also tightly constrained, whereas the free-\(H_0\) run remains strongly affected by the \(H_0\)--\(R_s\) degeneracy. The measured single-aperture stellar velocity dispersion of the main deflector is \(\sigma_{\rm LOS}=250^{+15}_{-21}\,{\rm km\,s^{-1}}\), based on ESO-MUSE spectroscopy \citep{sluse2019}. Our predicted dispersions are lower than this value, especially for the fixed-cosmology runs, but remain broadly consistent at the \(\sim2\sigma\) level given the observational uncertainty. This is not a direct tension because the calculation here is only a diagnostic: it is luminosity-weighted by the JWST MGE used in the lens model, which probes different stellar populations at different wavelengths than the VLT/MUSE measurement.

\section{Conclusions}

We have presented a composite strong-lensing analysis of the quadruply imaged quasar WFI2033--4723 using JWST/NIRCam imaging and the measured quasar time delays. Our goal is to study the baryonic and dark-matter structure of the lens; constraining the MST-like freedom with time-delay information and external cosmological assumptions is a prerequisite for this goal. We first construct an EPL baseline model to establish a reference lens model, and then replace the total mass profile with the star+gNFW framework. Our main conclusions are as follows.

\begin{enumerate}
\item The EPL baseline provides the reference Fermat-potential scale used in the composite analysis. The eight HMC chains converge to the same posterior despite being initialised from different starting points and sampling independent pixelized PSF corrections, with all reported \(\hat r\) values below 1.05. The inferred fpd are consistent with both the HST-based H0LiCOW analysis and the JWST-based TDCOSMO XX analysis, with statistical uncertainties that are broadly consistent with, and in some cases slightly smaller than, previous analyses. In our HMC analysis, the Matérn source-regularisation hyperparameters are sampled jointly with the lens model, so the uncertainty associated with the source regularization is marginalised within a single model specification. TDCOSMO XX explored a discrete set of modelling choices, including different source shapelet orders, mask sizes, PSF treatments, and flexion assumptions, and combined the resulting MCMC chains using BIC and kinematic weights. Their reported intervals therefore include an additional model-averaging component, especially if different modelling choices shift the fpd posterior and therefore increase the reported uncertainty.

\item Under both the Planck+DESI and Pantheon+SH0ES cosmological priors, the inferred stellar masses are \(M_\star=2.03^{+0.43}_{-0.62}\times10^{11}M_\odot\) and \(2.30^{+0.38}_{-0.43}\times10^{11}M_\odot\), respectively, placing the stellar normalisation between the Chabrier and Salpeter IMF expectations. This is similar to the intermediate IMF normalisation reported for NGC~6505 \citep{Euclid2025NGC6505}, and lighter than the Salpeter-like result for the Jackpot lens \citep{Tian2026}. The radial stellar \(M/L\) gradient is not strongly required by the data and remains degenerate with the stellar mass normalisation.

\item The dark-matter halo prefers a steep inner profile, with \(\gamma_{\rm in}=1.32^{+0.12}_{-0.15}\) for Planck+DESI and \(\gamma_{\rm in}=1.30^{+0.16}_{-0.21}\) for Pantheon+SH0ES. This is steeper than a standard NFW cusp and also steeper than the NFW-like halo inferred for the Jackpot lens using the same star+gNFW framework \citep{Tian2026}. It is also unusual compared with the near-NFW or mildly shallower population trends reported by Project DINOS II \citep{Sheu2025}, but broadly within the distribution from \citet{Oldham2018}.

\item We also tested whether the star+gNFW model can be used directly to infer \(H_0\). The cosmology-free run produces a converged posterior, but the inferred \(H_0\) is very sensitive to the inferred halo scale radius \(R_s\). In this chain, \(R_s\) is driven toward \(2^{\prime\prime}\). This is close to the lower edge of the adopted prior and likely unphysical. We therefore do not regard this \(H_0\) inference as reliable.
\end{enumerate}

The star+gNFW framework can fit the JWST image to the noise level and separate the stellar and dark-matter terms, but WFI2033--4723 alone does not constrain both the central profile and global halo scale radius well enough for an independent \(H_0\) measurement. Spatially resolved stellar kinematics will be needed to constrain the inner mass profile, break the remaining degeneracy with the stellar \(M/L\) gradient, and understand whether the very small \(R_s\) preferred by the cosmology-free lens model is physical or due to lens-modelling degeneracies.

\section*{Acknowledgements}
We gratefully acknowledge Martin Millon, Frédéric Courbin, Simon Birrer and the members of the TDCOSMO collaboration for insightful discussions that helped improve this work.

Numerical computations were done on the Sciama High Performance Compute (HPC) cluster, which is supported by the ICG, SEPNet, and the University of Portsmouth.

This work has received funding from the European Research Council (ERC) under the European Union's Horizon 2020 research and innovation program (LensEra: grant agreement No 945536). TC is funded by the Royal Society through a University Research Fellowship.

For the purpose of open access, the authors have applied a Creative Commons Attribution (CC BY) license to any Author Accepted Manuscript version arising.

\section*{Data Availability}
The JWST data used in this work are publicly available from the Mikulski Archive for Space Telescopes (MAST; \url{https://mast.stsci.edu/}).
The \texttt{Herculens} code is publicly available at \url{https://github.com/Herculens/herculens.git}.
The scripts used to generate the data products and figures are available at \url{https://github.com/astroskylee/Herculensedquasar/tree/main/WFI2033}.
The posterior samples and MCMC chains generated in this study are available from the corresponding author upon reasonable request.



\bibliographystyle{mnras}
\bibliography{example} 




\appendix

\section{Least-squares estimate of the PSF error map}
\label{app:psf_error_map}

The empirical PSF-error map used in the image likelihood is estimated from the residuals of the field-star PSF fits. For each star \(j\), let \(R_j(\boldsymbol{x})\) be the residual image in PSF-centred coordinates, \(M_j(\boldsymbol{x})\) the corresponding best-fitting model image, and \(\sigma_j(\boldsymbol{x})\) the expected noise map. We assume that, at each PSF coordinate \(\boldsymbol{x}\), the residual variance can be decomposed into the known random-noise contribution and an additional PSF-mismatch term whose amplitude scales with the local model flux, 
\begin{equation}
R_j^2(\boldsymbol{x})-\sigma_j^2(\boldsymbol{x}) \simeq \alpha_{\rm PSF}^2(\boldsymbol{x}) M_j^2(\boldsymbol{x}).
\end{equation}
This equation is fitted independently at each PSF pixel. Defining \(y_j=R_j^2-\sigma_j^2\), \(X_j=M_j^2\), and \(\beta=\alpha_{\rm PSF}^2\), the estimate is simply the zero-intercept least-squares solution obtained by minimizing
\begin{equation}
S(\beta)=\sum_j \left(y_j-\beta X_j\right)^2 .
\end{equation}
Setting \(\partial S/\partial \beta=0\) gives
\begin{equation}
\beta=\frac{\sum_j X_j y_j}{\sum_j X_j^2},
\end{equation}
and therefore
\begin{equation}
\alpha_{\rm PSF}^{2}(\boldsymbol{x})=\frac{\sum_j M_j^2(\boldsymbol{x})\left[R_j^2(\boldsymbol{x})-\sigma_j^2(\boldsymbol{x})\right]}{\sum_j M_j^4(\boldsymbol{x})}.
\end{equation}
The factors \(M_j^2\) and \(M_j^4\) therefore arise directly from fitting the excess residual variance against \(M_j^2\). This construction gives greater weight to bright, well-constrained PSF pixels and suppresses noisy estimates from low-flux regions, while retaining the interpretation of \(\alpha_{\rm PSF}(\boldsymbol{x})\) as a local fractional uncertainty in the PSF model.

\section{Additional posterior diagnostics}

This appendix collects supplementary posterior plots that support the discussion of the composite-model degeneracies and the cosmology-free \(H_0\) inference in the main text.

\begin{figure*}
    \centering
    \includegraphics[width=1\textwidth]{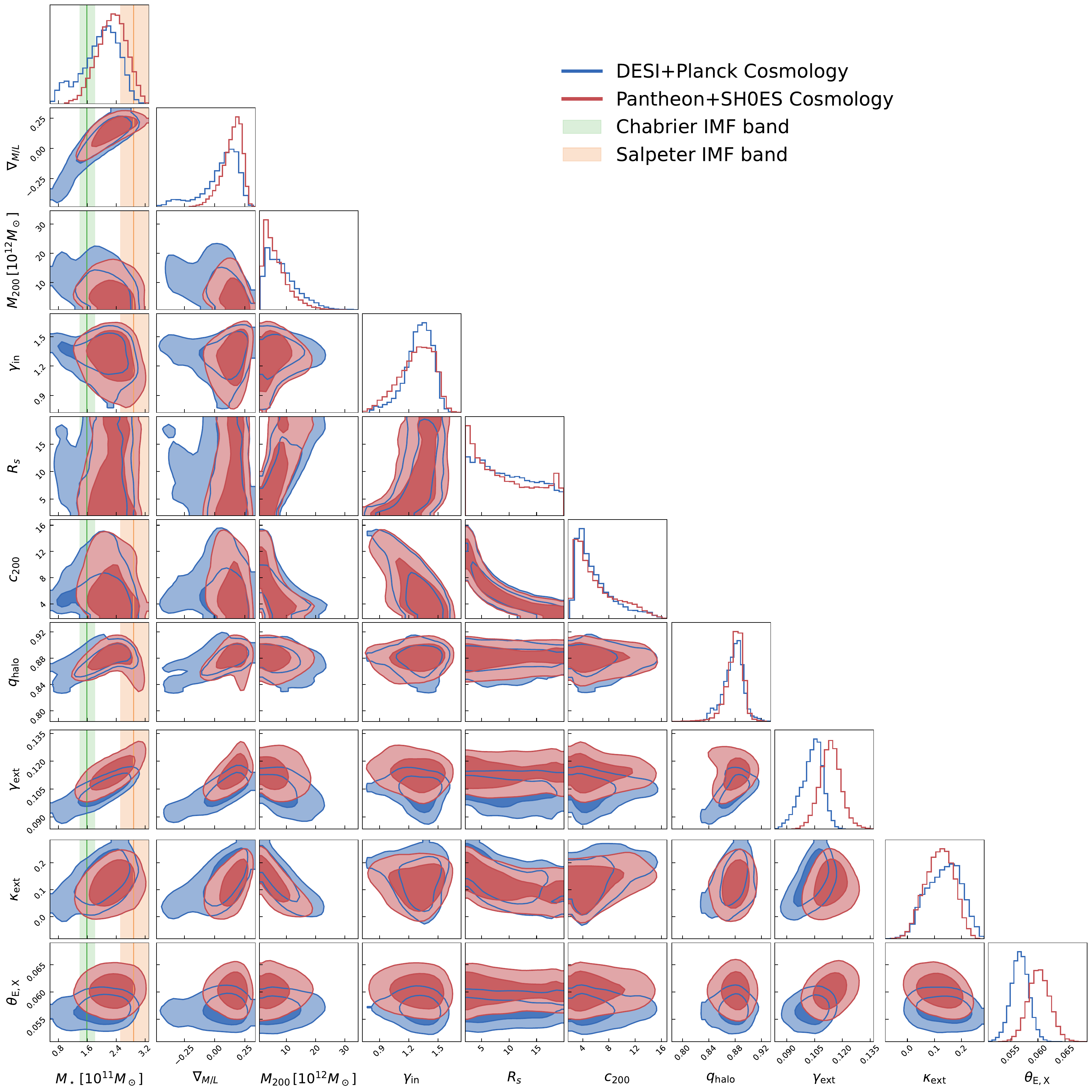}
    \caption{Posterior constraints on the derived stellar and halo parameters, including IMF-related quantities.}
    \label{fig:composite_imf_corner}
\end{figure*}

\begin{figure*}
    \centering
    \includegraphics[width=1\textwidth]{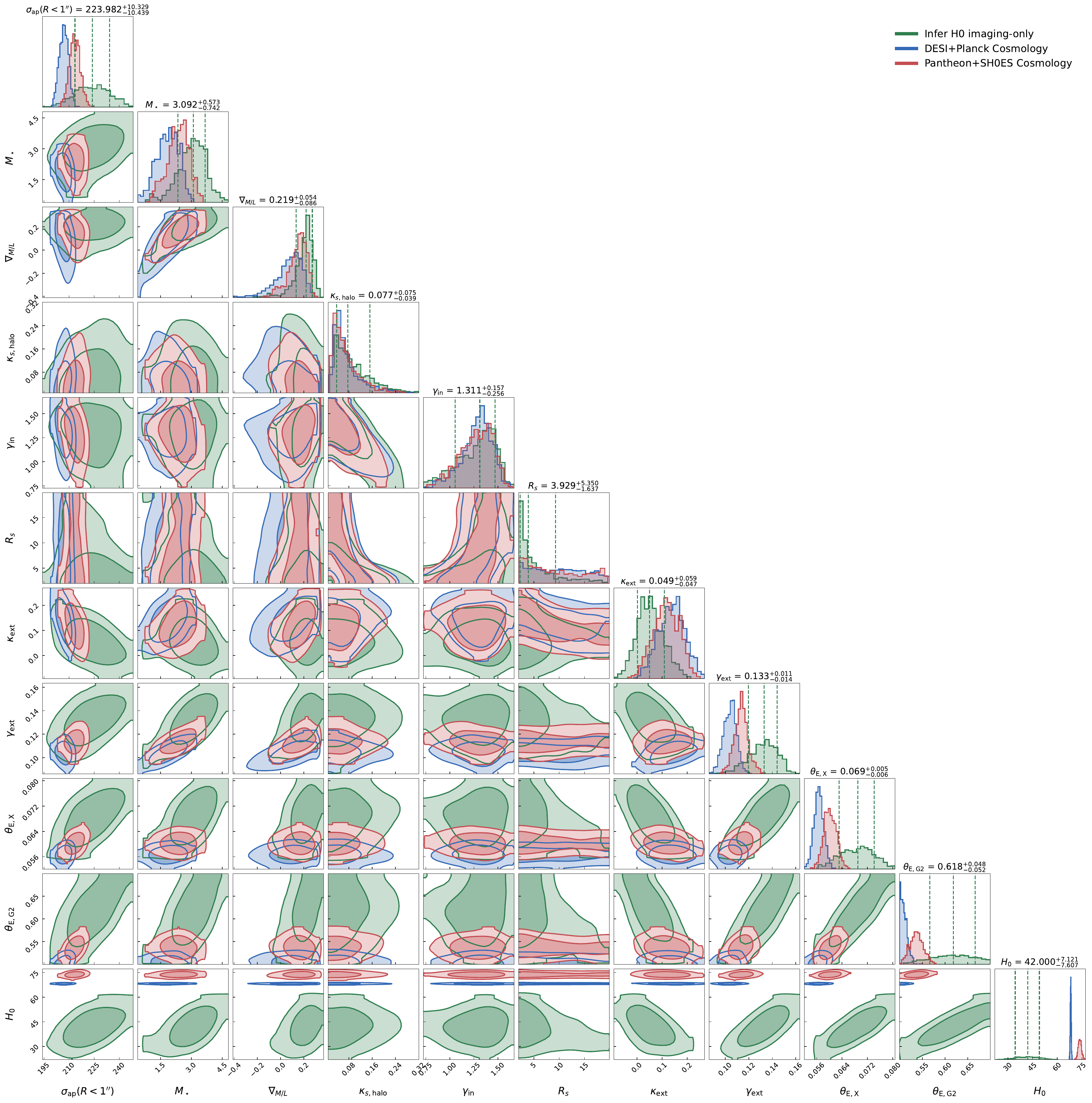}
    \caption{Posterior distribution for the composite model without an external cosmological prior, compared with the two fixed-cosmology composite models. The blue and red contours correspond to the same Planck+DESI and Pantheon+SH0ES models shown in Fig.~\ref{fig:composite_imf_corner}. The inferred \(H_0\) in the cosmology-free model is strongly degenerate with the dark-matter halo scale radius \(R_s\), which is driven toward the lower edge of the adopted prior.}
    \label{fig:jampy_sigma_ap}
\end{figure*}


\bsp	
\label{lastpage}
\end{document}